\documentclass[12pt]{article}

\usepackage[english]{babel}
\usepackage[latin1]{inputenc}
\usepackage{epsfig}
\usepackage{color}
\usepackage{rotating}
\usepackage{psfrag}

\usepackage[intlimits]{amsmath}
\usepackage{amssymb}
\usepackage{slashed}
\usepackage{feynmp}

\usepackage{url}
\usepackage{graphicx}

\numberwithin{equation}{section}

\long\def\symbolfootnote[#1]#2{\begingroup\def\thefootnote{\fnsymbol{footnote}}
\footnote[#1]{#2}\endgroup}

\def\lsim{\mathrel{\raise.3ex\hbox{$<$\kern-.75em\lower1ex\hbox{$\sim$}}}}
\def\gsim{\mathrel{\raise.3ex\hbox{$>$\kern-.75em\lower1ex\hbox{$\sim$}}}}

\renewcommand{\Re}{\mbox{Re}}

\newcommand{\sdm}{{\phi_\text{DM}}} 
\newcommand{\fdm}{{\psi_\text{DM}}} 
\newcommand{\nn}{N} 
\newcommand{\med}{\Sigma} 
\newcommand{\medV}{V} 

\newcommand{\lamFL}{\lambda_{q\psi}^L} 

\newcommand{\lamFR}{\lambda_{q\psi}^R} 
\newcommand{\lamNNL}{\lambda_{q\nn}^L} 
\newcommand{\lamNNLast}{\lambda_{q\nn}^{L \ast}}
\newcommand{\lamNNR}{\lambda_{q\nn}^R} 
\newcommand{\lamNNRast}{\lambda_{q\nn}^{R \ast}}

\newcommand{\GeV}{\,{\rm GeV}}

\begin{document}

\setlength{\unitlength}{1mm}

\date{\mbox{ }}

\title{
{\normalsize
TUM-HEP 840/12\hfill\mbox{}\\
DESY 12-084\hfill\mbox{}\\
FTPI-MINN-12/18\hfill\mbox{}\\
UMN-TH-3104/12\hfill\mbox{}\\}
\vspace{1.5cm} 
\bf Constraints on Hadronically Decaying\\Dark Matter\\[8mm]}

\author{Mathias Garny$^{1,2}$, Alejandro Ibarra$^1$, David Tran$^{1,3}$ \\[2mm]
{\normalsize\it $^1$Physik-Department T30d, Technische Universit\"at M\"unchen,}\\[-0.05cm]
{\it\normalsize James-Franck-Stra\ss{}e, 85748 Garching, Germany}\\[1.5mm]
{\normalsize\it $^2$Deutsches Elektronen-Synchrotron DESY,}\\[-0.05cm]
{\it\normalsize Notkestra\ss{}e 85, 22603 Hamburg, Germany}\\[1.5mm]
{\normalsize\it $^3$School of Physics and Astronomy, University of Minnesota,}\\[-0.05cm]
{\it\normalsize 116 Church Street SE, Minneapolis, MN 55455, USA}}
\maketitle

\thispagestyle{empty}

\begin{abstract}
\noindent
We present general constraints on dark matter stability in hadronic decay channels derived from measurements of cosmic-ray antiprotons. We analyze various hadronic decay modes in a model-independent manner by examining the lowest-order decays allowed by gauge and Lorentz invariance for scalar and fermionic dark matter particles and present the corresponding lower bounds on the partial decay lifetimes in those channels. We also investigate the complementarity between hadronic and gamma-ray constraints derived from searches for monochromatic lines in the sky, which can be produced at the quantum level if the dark matter decays into quark--antiquark pairs at leading order.
\end{abstract}

\newpage

\section{Introduction}

Despite the overwhelming evidence for the existence of dark matter in our Universe, very little is known about its properties from the point of view of particle physics. One of the most striking properties of the particles comprising the dark matter is their longevity, evidenced by their survival from production in the early Universe to the present day. Among the hundreds of matter particles which have been produced and studied over the last century at accelerators, only very few have lifetimes on a cosmological scale: the proton, the electron and the three neutrinos. The longevity of the electron, which is the lightest charged particle, can be attributed to the conservation of electric charge. The longevity of the lightest neutrino, which is also the lightest fermion, is due to Lorentz symmetry. The longevity of the other two neutrinos, which can decay into the lightest neutrino and a photon, is because of their tiny masses. Lastly, the longevity of the proton can be attributed to the accidental conservation of baryon number in the renormalizable Standard Model Lagrangian, which might be broken by higher-dimensional operators. However, no general principle is known which would necessitate a cosmologically long dark matter lifetime. On the contrary, the longevity of the dark matter particle can only be understood within concrete models of dark matter, and is usually due to additional symmetries imposed from the start.

In the absence of a fundamental reason to guarantee the absolute stability of the dark matter particle, it is important to analyze the potential signatures of the decay of dark matter particles and the constraints on the dark matter lifetime which follow from cosmic-ray measurements. Generally, indirect constraints on dark matter stability can be inferred from observations of gamma rays~\cite{Ibarra:2009nw,Cirelli:2009dv,Chen:2009uq,Hutsi:2010ai,Dugger:2010ys,Cirelli:2012ut}, neutrinos~\cite{Hisano:2008ah,Covi:2009xn,Buckley:2009kw,Esmaili:2012us}, electrons/positrons~\cite{Cirelli:2008pk,Nardi:2008ix,Ibarra:2008jk,Ibarra:2009dr,Ishiwata:2009vx} and antiprotons/antideuterons~\cite{Cholis:2010xb,Evoli:2011id,Ibarra:2009tn,Kadastik:2009ts,Cui:2010ud}. In this paper we focus on the hadronic decay modes of general dark matter candidates. We use the measurements of the antiproton-to-proton fraction performed with the PAMELA satellite instrument~\cite{Adriani:2010rc,Adriani:2008zq} to constrain the minimum dark matter lifetime in hadronic decay modes, examining a number of possibilities for the coupling of the dark matter particle to the Standard Model particles. We also argue that the partial lifetime for the decay of dark matter particles into a quark--antiquark pair and a neutral fermion, $\psi_\text{DM} \to q \bar{q} N$, can also be constrained using gamma-ray observations; more concretely via searches for monoenergetic gamma-rays which can be generated at the quantum level by radiating photons from virtual quarks. Instead of photons, weak gauge bosons can also be generated at next-to-leading order in this manner, even in models where the dark matter does not decay hadronically at leading order.

This paper is organized as follows. In Section~\ref{sec:antiproton_production} we review the production and propagation of antiprotons in the Galaxy. We then derive constraints on the partial dark matter lifetime in various hadronic decay modes in Section~\ref{sec:tree_level}. In Section~\ref{sec:gamma_rays} we discuss the generation of monochromatic gamma rays from decays into quarks and compare the resulting constraints with those from cosmic-ray antiprotons. The production of antiprotons from radiative decays into weak gauge bosons is analyzed in Section~\ref{sec:weak_gauge_bosons}. We present our conclusions in Section~\ref{sec:conclusions}.

\section{Antiproton Production from Dark Matter Decay}
\label{sec:antiproton_production}

Antiprotons are a rare component of the cosmic radiation (at 10 GeV kinetic energy about one antiproton is observed for every $10^4$ protons~\cite{Adriani:2010rc}). These cosmic antiprotons are believed to be overwhelmingly or exclusively of secondary origin by spallation of primary cosmic rays on interstellar gas, a hypothesis which is in good quantitative agreement with observations~\cite{Bringmann:2006im}. Therefore, the cosmic-ray antiproton flux constitutes a sensitive probe for particle production in the Galaxy from exotic sources like dark matter decay. In this section we briefly review the production and propagation in the Galaxy of antiprotons from dark matter decay.

Given a population of dark matter particles with mass $m_\text{DM}$ and lifetime $\tau_\text{DM}$, the production rate of antiprotons per unit energy and unit volume at a 
position $\vec{r}$ with respect to the center of the Galaxy is given by
\begin{equation}
  Q_{\bar{p}}(E,\vec{r})=\frac{\rho_\text{DM}(\vec{r})}{m_{\rm DM}\,\tau_{\rm DM}}\frac{dN_{\bar{p}}}{dE}\;,
  \label{eqn:source-term}
\end{equation}
where $dN_{\bar p}/dE$ is the energy spectrum of antiprotons produced in the decay and
$\rho_\text{DM}(\vec{r})$ is the density profile of the distribution of dark matter in the Milky Way halo. For definiteness we will adopt the spherically symmetric
Navarro-Frenk-White halo density profile~\cite{Navarro:1995iw,Navarro:1996gj}:
\begin{equation}
  \rho_\text{DM}(r)=\frac{\rho_0}{(r/r_\text{c})
  [1+(r/r_\text{c})]^2}\;,
\end{equation}
with $\rho_0\simeq 0.26\,{\rm GeV}/{\rm cm}^3$ and $r_c\simeq 20 ~\rm{kpc}$,
which corresponds to a local dark matter density of $\rho_\odot= 0.4~\text{GeV}/\text{cm}^3$~\cite{Catena:2009mf,Weber:2009pt,Salucci:2010qr}. Due to the linear dependence on the dark matter density, different choices for the halo profile yield very similar constraints~\cite{Ibarra:2008qg}.

After being produced in the Milky Way halo, antiprotons propagate through the Galaxy and its diffusive magnetic halo in a rather complicated manner before reaching the Earth. Cosmic-ray propagation in the Galaxy is commonly described using a stationary two-zone diffusion model with cylindrical boundary conditions~\cite{ACR}. Due to the relatively large antiproton mass, we can neglect energy losses as well as reacceleration effects in our treatment~\cite{Maurin:2002ua}. We furthermore neglect catastrophic losses from antiproton annihilation with ordinary matter in the Galactic disk since they have a negligible effect on the total antiproton flux. Under these approximations, the number density $f_{\bar p}(T,\vec{r},t)$ of antiprotons per unit kinetic energy $T$ satisfies the following transport equation:
\begin{equation}
  0=\frac{\partial f_{\bar p}}{\partial t}=
  \vec\nabla \cdot [K(T,\vec{r})\vec\nabla f_{\bar p}-
  \vec{V_c}(\vec{r})  f_{\bar p}] +Q_{\bar p}(T,\vec{r})\;.
  \label{transport}
\end{equation}

The first term on the right-hand side of the transport equation is the
diffusion term, which accounts for the propagation through the tangled
Galactic magnetic field. The diffusion coefficient $K(T,\vec{r})$ is assumed
to be constant throughout the diffusion zone and is parametrized by $K(T)=K_0
\,\beta\, {\cal R}^\delta$, where $\beta=v/c$ is the antiproton speed as a fraction of the speed of light and ${\cal R}$ is the rigidity
of the particle, which is defined as the momentum in GeV per unit charge,
${\cal R}\equiv p({\rm GeV})/|Z|$. The second term is the convection term, 
which accounts for the drift of antiprotons away from the disk induced 
by the Milky Way's Galactic wind. Following Ref.~\cite{MDT+01} we will assume 
that it has axial direction and that it is constant inside the diffusion region above and below the Galactic disk: $\vec{V}_\text{c}(\vec{r})=V_\text{c}\; {\rm sign}(z)\; \vec{k}$. Lastly, $Q_{\bar{p}}(T,\vec{r})$ is the source term of cosmic antiprotons as defined in Eq.~(\ref{eqn:source-term}). 

The solution of the transport equation at the heliospheric boundary at our position in the Galaxy, $r = r_\odot$, $z = 0$, can be formally expressed by the convolution
\begin{equation}
  f_{\bar p}(T)=\frac{1}{m_{\rm DM} \tau_{\rm DM}}
  \int_0^{T_{\rm max}}dT^\prime\; G_{\bar p}(T,T^\prime) \;
  \frac{dN_{\bar p}(T^\prime)}{dT^\prime}\;,
  \label{eqn:fconv}
\end{equation}
where $T_{\rm max} = m_{\rm DM} - m_p$, with $m_p$ being the proton mass. Analytical
and numerical expressions for the Green's function $G_{\bar p}(T,T^\prime)$ derived within the semi-analytical framework described above can be found in~\cite{Ibarra:2008qg}.

Given the number density, the flux of primary antiprotons from dark matter decay at the heliospheric boundary is finally given by
\begin{equation}
  \Phi_{\bar p}^{\rm{DM}}(T) = \frac{v(T)}{4 \pi} f_{\bar p}(T)\;,
  \label{flux-antiproton}
\end{equation}
where $v$ is the speed of the antiprotons as a function of their kinetic energy. 

The computation of the local antiproton flux suffers from various sources of uncertainty. The most important one stems from the choice of the propagation parameters that enter in the transport equation (the normalization of the diffusion coefficient, the power-law index of the rigidity, the height of the diffusion zone and the speed of the convective wind). This uncertainty in the model parameters amounts to as much as two orders of magnitude. In view of this large uncertainty we calculate our constraints for the full range of parameters compatible with observations. In contrast, the uncertainties on the choice of the dark matter halo profile and the local dark matter density are subdominant, and amount at most to a factor of order one~\cite{Ibarra:2008qg}. Due to the linear dependence of the fluxes on the dark matter density, our constraints can easily be rescaled to other values. Therefore, we will just present results for the Navarro-Frenk-White halo profile with $\rho_\odot= 0.4~\text{GeV}/\text{cm}^3$, whereas we will present the entire range of results for three different propagation models that are consistent with the observed boron-to-carbon (B/C) ratio and that give the maximum (MAX), medium (MED) and minimum (MIN) antiproton flux~\cite{MDT+01}. The relevant parameters are summarized in Tab.~\ref{tab:param-antiproton}.

\begin{table}[t]
  \begin{center}
    \begin{tabular}{|c|cccc|}
      \hline
      Model & $\delta$ & $K_0\,({\rm kpc}^2/{\rm Myr})$ & $L\,({\rm kpc})$
      & $V_c\,({\rm km}/{\rm s})$ \\
      \hline 
      MIN & 0.85 & 0.0016 & 1 & 13.5 \\
      MED & 0.70 & 0.0112 & 4 & 12 \\
      MAX & 0.46 & 0.0765 & 15 & 5 \\
      \hline
    \end{tabular}
    \caption{\label{tab:param-antiproton} \small Astrophysical parameters
    compatible with the observed B/C ratio that yield the minimum (MIN), 
    medium (MED)
    and maximum (MAX) flux of antiprotons~\cite{MDT+01}.}
  \end{center}
\end{table}

Lastly, antiproton propagation inside the Solar System is influenced by solar
modulation effects, which have an effect on the locally observed low-energy 
antiproton spectrum. Under the force field
approximation~\cite{solar-modulation}, the antiproton flux at the top of the
Earth's atmosphere is related to the interstellar antiproton flux~\cite{perko}
by the simple relation:
\begin{equation}
  \Phi_{\bar p}^{\rm TOA}(T_{\rm TOA})=
  \left(
  \frac{2 m_p T_{\rm TOA}+T_{\rm TOA}^2}{2 m_p T_{\rm IS}+T_{\rm IS}^2}
  \right)\,
  \Phi_{\bar p}^{\rm IS}(T_{\rm IS}),
\end{equation}
where $T_{\rm IS}=T_{\rm TOA}+\phi_F$, with $T_{\rm IS}$ and $T_{\rm TOA}$
being the antiproton kinetic energies at the heliospheric boundary and at the
top of the Earth's atmosphere, respectively, and $\phi_F$ being the solar
modulation parameter, which varies between 500 MV and 1.3 GV over the
eleven-year solar cycle. In order to compare our predictions with the PAMELA data we will take $\phi_F=550$ MV~\cite{Bar97}. 

\section{Tree-Level Decays}
\label{sec:tree_level}
We begin our study of the constraints on the dark matter lifetime by examining the tree-level decays of scalar and fermionic dark matter particles into weak gauge bosons, Higgs bosons and quarks, which subsequently hadronize, producing antiprotons and other particles.

\subsection{Decay of Spin-1/2 Dark Matter Particles}
\label{spin-half-tree}

We first consider the decays of a spin-1/2 dark matter particle. In this case,
the simplest decays allowed by Lorentz invariance are 
two-body decays into a spin-1/2 fermion 
and a boson (scalar or vector) and three-body decays into fermions.
Decays with larger multiplicities in the final state, such as three-body
decays into a spin-1/2 fermion and two scalars, four-body decays into
three fermions and one gauge boson or five-body decays into fermions will
not be considered in this paper. To keep the analysis as general as possible
we will not specify the nature of the neutral fermion in the final state,
which we denote by $N$. This particle could be a neutrino or a heavier 
particle, 
possibly another dark matter component, such as a gravitino or a gaugino of 
some hidden-sector gauge group.

The lowest order decay modes of a spin-1/2 dark matter particle are:
\begin{itemize}
\item Decay into a spin-1/2 fermion and a vector boson: 
$\psi_\text{DM} \to Z^0 N$, $\psi_\text{DM} \to W^\pm\ell^\mp$
\item Decay into a spin-1/2 fermion and a Higgs boson: 
$\psi_\text{DM} \to h^0 N$
\item Decay into two quarks and a spin-1/2 fermion: $\psi \to u_i \bar u_j N$,
$\psi_\text{DM} \to d_i \bar d_j N$,
$\psi_\text{DM} \to u_i \bar d_j \ell$
\end{itemize}
Here, $i,j=1,2,3$ are quark generation indices (note that the lepton 
generation is irrelevant for antiproton production). In the following we only regard the case that the two quarks are of the same flavor, $i = j$.

The production of antiprotons by fragmentation of the weak gauge bosons, the Higgs boson and the quarks was simulated using the event generator PYTHIA 6.4~\cite{Sjostrand:2006za}. Taking these spectra as input in the source term in Eq.~\ref{eqn:source-term} we then calculate the antiproton-to-proton fraction at Earth by solving the transport equation, Eq.~(\ref{transport}), for the MIN, MED and MAX set of propagation parameters that we list in Table \ref{tab:param-antiproton}. From the resulting antiproton-to-proton ratios we calculate lower bounds on the dark matter lifetime for dark matter masses in the range 100 GeV -- 20 TeV by comparing the results to the PAMELA measurements of the antiproton-to-proton fraction. We have also adopted the background flux of secondary antiprotons produced in spallations of cosmic rays in the interstellar medium calculated in~\cite{Bringmann:2006im} which was calculated within the same two-zone diffusion model and agrees well with the measured antiproton-to-proton ratio. We derive conservative constraints on the partial dark matter lifetime from the requirement that the total antiproton-to-proton ratio should not exceed the PAMELA measurements at 95\% C.L. as determined by the $\chi^2$ statistic
\begin{equation}
\chi^2 = \sum_i \frac{[\Phi_\text{tot}^{\bar{p}}(T_i)/\Phi_\text{tot}^p(T_i) - x_i]^2}{\sigma_i^2} \;,
\end{equation}
where the $x_i$ are the measured values of the antiproton-to-proton ratio at kinetic energies $T_i$, and the $\sigma_i$ are the corresponding statistical errors. The $\Phi(T_i)$ denote the top-of-atmosphere fluxes of protons and antiprotons. The total flux $\Phi_\text{tot}(T) = \Phi_\text{DM}(T) + \Phi_\text{bkg}(T)$ is the sum of contributions from dark matter decay and the ``background'' from ordinary astrophysical production by cosmic-ray spallation. We also regard the case of a pure antiproton signal from dark matter decay, where we disregard the astrophysical background, and for which we use the following ``$\chi^2$'' statistic,
\begin{equation}
\chi^2 = \sum_i \frac{\{R[\Phi_\text{DM}^{\bar{p}}(T_i)/\Phi_\text{tot}^p(T_i) - x_i]\}^2}{\sigma_i^2} \;,
\end{equation}
where $R(x)=x$ for $x>0$ and $R(x)=0$ for $x\leq 0$, so that only fluxes that exceed the measurements are penalized when we regard the case of a pure dark matter-induced signal with no background.

Given that astrophysical models of secondary antiproton production by cosmic-ray spallation can quantitatively reproduce the observed antiproton-to-proton fraction well, the constraints derived in the former analysis are significantly more stringent than in the latter.

\subsubsection{Two-Body Decays}
We analyze the decay of a spin-1/2 dark matter particle in a simple toy model where we assume a Lagrangian that describes the interaction between dark matter, weak gauge bosons and fermions, where we assume the dark matter to be a Majorana particle. The two-body decay of a spin-1/2 dark matter particle with mass $m_{\rm DM}$ into a weak gauge boson and a fermion is induced by the following terms in the Lagrangian,
\begin{equation}
-\mathcal{L} = \bar{\psi}_\text{DM} \gamma^\mu \left[g_{\psi Z}^L P_L + g_{\psi Z}^R P_R\right] N \, Z_\mu+ \text{h.c.}\;,
\end{equation}
\begin{equation}
-\mathcal{L}= \bar{\psi}_\text{DM} \gamma^\mu \left[g_{\psi W}^L P_L + g_{\psi W}^R P_R\right] \ell \, W_\mu+ \text{h.c.}\;,
\end{equation}
where $P_{L,R} = \frac{1}{2} (1 \mp \gamma^5)$ are the left and right-handed chirality projectors and the $g_{ij}^{L/R}$ are coupling constants.
These interaction terms yield the following decays widths for the two-body decays $\psi_\text{DM} \to Z^0 N$ and $\psi_\text{DM} \to W^\pm \ell^\mp$,
\begin{align}
\Gamma_{Z} = &\frac{m_{\rm DM}}{32\pi} \sqrt{\left(1-\frac{(m_\nn+M_Z)^2}{m_{\rm DM}^2}\right)\left(1-\frac{(m_\nn-M_Z)^2}{m_{\rm DM}^2}\right)}\times\Bigg\{\bigg[1+\frac{m_\nn^2-2M_Z^2}{m_{\rm DM}^2}\nonumber\\
&+\frac{m_{\rm DM}^2}{M_Z^2}
\left(1-\frac{m_\nn^2}{m_{\rm DM}^2}\right)^2\bigg]\Big(|g_{\psi Z}^L|^2+|g_{\psi Z}^R|^2\Big)-\frac{12m_\nn}{m_{\rm DM}}\Re\Big(g_{\psi Z}^Lg_{\psi Z}^{R*}\Big)\Bigg\} \;, 
\end{align}
and
\begin{align}
\Gamma_{W} = &\frac{m_{\rm DM}}{16\pi } \left(1+\frac{m_{\rm DM}^2}{2M_W^2}\right)\left(1-\frac{M_W^2}{m_{\rm DM}^2}\right)^2 \Big(|g_{\psi W}^L|^2+|g_{\psi W}^R|^2\Big)\,,
\end{align}
respectively. From these decay widths we see that the couplings must be very strongly suppressed in order to yield dark matter lifetimes that exceed the age of the Universe.

The decay of dark matter particles produces monoenergetic weak gauge bosons with an energy 
given by
\begin{equation}
E_Z = \frac{m_{\rm DM}}{2}\left(1+\frac{M_Z^2-m_\nn^2}{m_{\rm DM}^2}\right) \,,
\end{equation}
and
\begin{equation}
E_W = \frac{m_{\rm DM}}{2}\left(1+\frac{M_W^2}{m_{\rm DM}^2}\right) \,,
\end{equation}
respectively, where the charged lepton mass has been neglected.

\begin{figure}[h!]
\begin{center}
\psfrag{mdmhhhhhhhhhhh}{\scriptsize $m_\text{DM}$ [GeV]}
\psfrag{gammadmhhhh}{\scriptsize \hspace{0.2cm} $\Gamma^{-1}$ [s]}
\includegraphics[width=80mm]{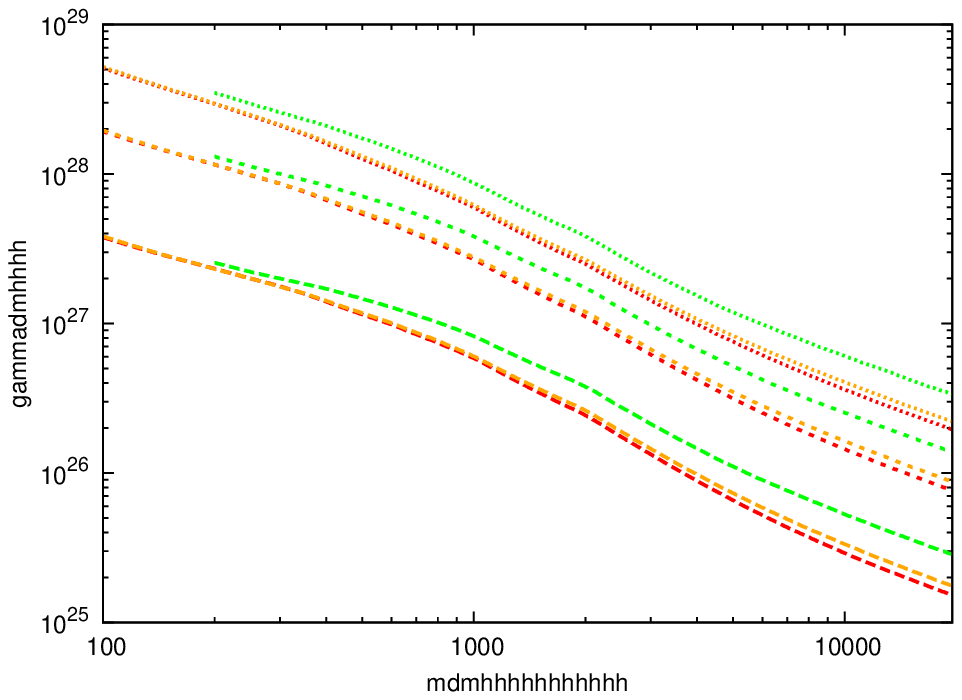}~~~\includegraphics[width=80mm]{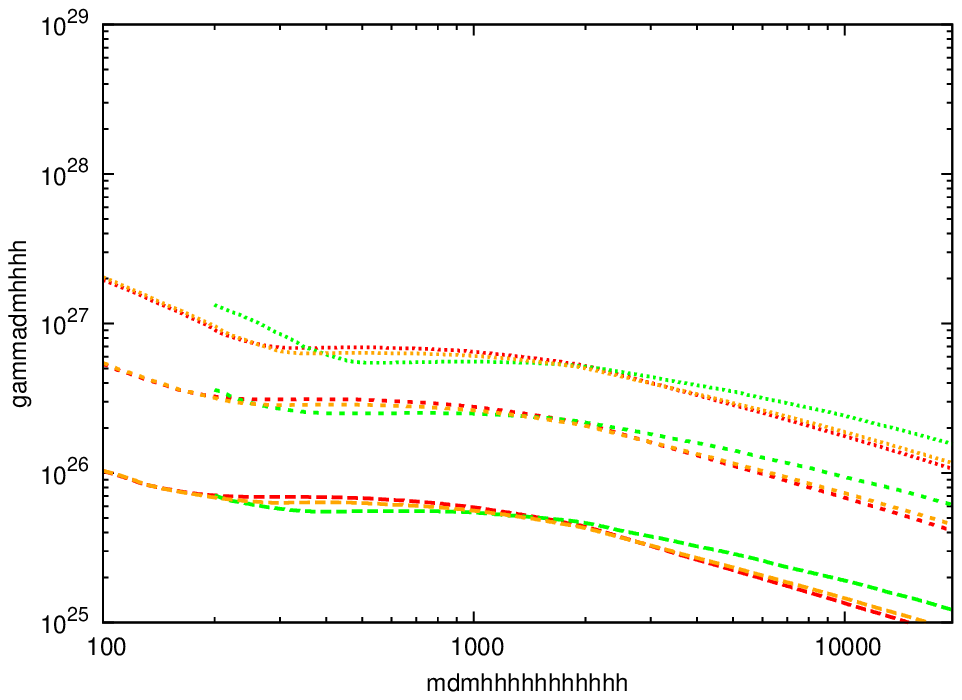}
\end{center}
\caption{Lower limits on the partial  lifetime of a fermionic dark matter particle which decays into a lepton and a gauge or Higgs boson from the requirement of not exceeding the antiproton-to-proton fraction as measured by PAMELA at 95\% C.L. The red line corresponds to the decay $\psi_\text{DM} \to W^\pm \ell^\mp$, the orange line to $\psi_\text{DM} \to Z^0 \nu$ and the green line to  $\psi_\text{DM} \to h^0 \nu$, while the long-dashed, short-dashed and dotted lines correspond to the MIN, MED and MAX propagation model parameters, respectively (see text for details). \textit{Left panel:} Constraints for primary antiprotons from dark matter decay + secondary antiprotons from cosmic-ray spallation. \textit{Right panel:} Constraints for primary antiprotons from dark matter decay only.}
\label{fig:spin12_bosons}
\end{figure}

We show in Fig.~\ref{fig:spin12_bosons}, in red lines, the constraints on the partial dark matter lifetime as a function of the dark matter mass in the case that the dark matter particle decays with 100\% branching ratio via $\psi_\text{DM} \to W^\pm \ell^\mp$. The long-dashed, short-dashed and dotted lines correspond, respectively, to the MIN, MED and MAX propagation parameters shown in Table~\ref{tab:param-antiproton}. The results for the decay $\psi_\text{DM} \to Z^0 \nu$ are similar to the case $\psi_\text{DM} \to W^\pm \ell^\mp$, shown in Fig.~\ref{fig:spin12_bosons} as orange lines, except in the region of dark matter masses which are close to the gauge boson mass. It follows from our calculation that the PAMELA measurements of the antiproton-to-proton fraction require, for dark matter particles which decay into weak gauge bosons, a corresponding partial lifetime longer than ${\cal O}(10^{26}-10^{27})$ s at $m_{\rm DM}= 1$ TeV, depending on the propagation model parameters when using the most conservative approach, and  ${\cal O}(10^{27}-10^{28})$ when using the more stringent and realistic approach which takes into account the background.

Similarly, the two-body decay of a spin-1/2 dark matter fermion with mass $m_{\rm DM}$ into the Standard Model Higgs boson and a neutral particle is induced by the following interaction term in the Lagrangian,
\begin{equation}
-{\cal L} = \bar{\psi}_\text{DM} [\lambda_{\psi h}^L P_L + \lambda_{\psi h}^R P_R] N h + \text{h.c.}\;,
\end{equation}
where the $\lambda_{ij}^{L/R}$ are coupling constants. This interaction yields a decay width given by
\begin{align}
\Gamma_h = &\frac{m_{\rm DM}}{32\pi}\sqrt{\left(1-\frac{(m_\nn+m_h)^2}{m_{\rm DM}^2}\right)\left(1-\frac{(m_\nn-m_h)^2}{m_{\rm DM}^2}\right)}\nonumber\\
&\times \left\{
\left(1-\frac{m_h^2-m_\nn^2}{m_{\rm DM}^2}\right)\Big(|\lambda_{\psi h}^L|^2+|\lambda_{\psi h}^R|^2\Big)+\frac{4 m_\nn}{m_{\rm DM}}\Re\Big(\lambda_{\psi h}^L\lambda_{\psi h}^{R*}\Big)
\right\}\,,
\end{align}
while the Higgs boson produced in the decay has an energy of
\begin{eqnarray}
E_h &=& \frac{m_{\rm DM}}{2}\left(1+\frac{m_h^2-m_\nn^2}{m_{\rm DM}^2}\right) \,.
\end{eqnarray}

In Fig.~\ref{fig:spin12_bosons}, green line, we show the lower bound on the dark matter lifetime in the decay channel $\psi_\text{DM} \to h^0 \nu$ as a function of the dark matter mass, where we assume that the Higgs boson is Standard Model-like and has a mass of $m_{h^0} = 125 \GeV.$

\subsubsection{Three-Body Decays}

\begin{figure}[t]
  \begin{center}
    \includegraphics{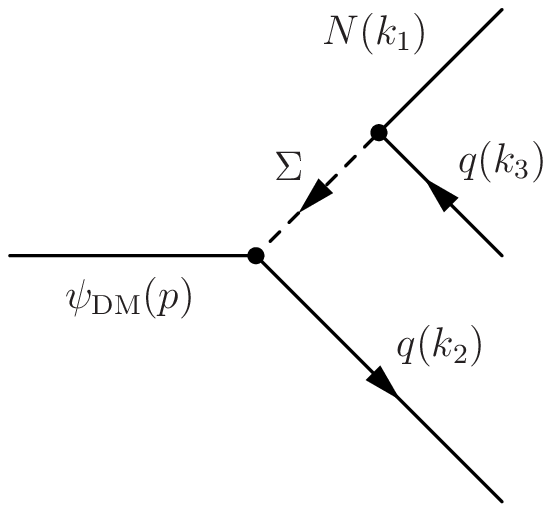}
    ~~~~~
    \includegraphics{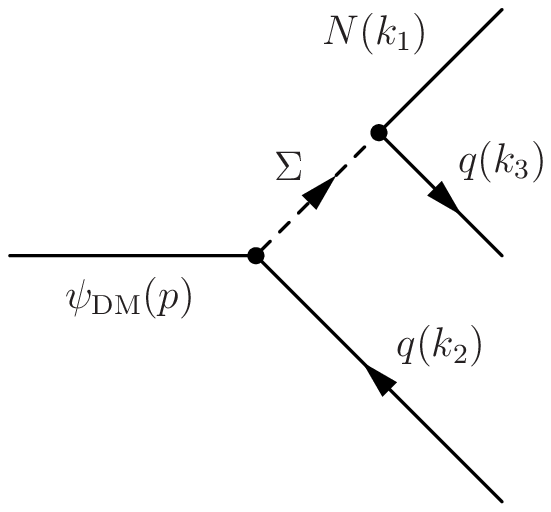}\\
    \caption{Tree-level diagrams contributing to the three-body decay
    $\psi_\text{DM} \to q\bar q N$ of fermionic Majorana dark matter,
    mediated by a heavy charged scalar $\Sigma$. Instead of the intermediate
    scalar $\Sigma$, the decay can also be mediated by a vector $V$.}
    \label{fermion_tree}
  \end{center}
\end{figure}

The energy spectrum of the antiprotons produced in three-body decays of dark matter particles contains some model dependence. In this paper we consider two possibilities, depending on whether the decay is mediated by a virtual charged scalar or a virtual charged vector particle, where we assume these particles to be much heavier than the dark matter mass. A generic interaction Lagrangian of the dark matter particle with a heavy charged scalar $\Sigma$ which induces the decay $\fdm \to q\bar q N$ is
\begin{eqnarray}\label{Leffscalar}
   \mathcal{L}_\text{eff} & = & - \bar\psi_\text{DM} \left[ \lamFL  P_L + \lamFR  P_R \right] q \, \med^\dag   - \bar\nn \left[ \lamNNL  P_L + \lamNNR  P_R \right] q \, \med^\dag
                       + \mbox{h.c.}
\end{eqnarray}
In the limit $m_q\ll m_{\rm DM} \ll m_\Sigma$  the partial decay width for the decay $\psi_\text{DM} \to q  \bar q  N$
is given by
\begin{equation}\label{eqn:fermion_3body_width}
  \Gamma(\psi_\text{DM} \to q \, \bar q \, N) = 
  \frac{m_{\text{DM}}^5}{128(2 \pi)^3 m_\Sigma^4} \left\{C_1^\Sigma
  F_1(m_N^2/m_{\text{DM}}^2) + C_2^\Sigma
  F_2(m_N^2/m_{\text{DM}}^2)\right\}.
\end{equation}
The constants $C_1^\Sigma$, $C_2^\Sigma$ are determined by the couplings as
\begin{align}
  C_1^\Sigma &\equiv \left(|\lambda_{q \psi}^L|^2 + |\lambda_{q
  \psi}^R|^2\right) \left(|\lambda_{q N}^L|^2 + |\lambda_{q N}^R|^2\right) 
  - \eta\,\text{Re} \left(\lambda_{q \psi}^L \lambda_{q N}^{L*} 
  \lambda_{q \psi}^R \lambda_{q  N}^{R*}\right),\\
  C_2^\Sigma &\equiv 2 \eta\,
  \text{Re}\left[\left(\lambda_{q \psi}^L \lambda_{q N}^{L*}\right)^2 +
  \left(\lambda_{q \psi}^R \lambda_{q N}^{R*}\right)^2\right].
\end{align}
Here, $\eta \equiv \eta_{\psi_\text{DM}} \eta_N = \pm 1$ depending on the
\textsl{CP} eigenvalues of $\psi_\text{DM}$ and $N$. The kinematical
functions, on the other hand, are given by
\begin{align}
  F_1(x) &\equiv (1 - x^2)(1 + x^2 - 8x) - 12x^2 \ln(x), \label{eqn:F1}\\
  F_2(x) &\equiv \sqrt{x} [(1 - x) (1 + 10x + x^2) + 6x (1 + x) \ln(x)].
  \label{eqn:F2}
\end{align}
In the hierarchical limit $m_N/m_{\text{DM}} \to 0$, the
kinematical functions satisfy
\begin{equation}
\label{Fs-hierarchical}
  F_1(x) \simeq 1, ~~~ F_2(x) \simeq \sqrt{x}\,,
\end{equation}
whereas in the degenerate limit $m_N/m_{\text{DM}} \to 1$, one
gets
\begin{equation}
\label{Fs-degenerate}
  F_1(x) \simeq \frac{2}{5}(1 - x)^5, ~~~ F_2(x) \simeq \frac{1}{10}(1 - x)^5\,.
\end{equation}

In the case of a vector interaction, on the other hand, we consider an effective Lagrangian of the form
\begin{equation}
\mathcal{L}_\text{eff}^V = -\bar{\psi}_\text{DM} \gamma^\mu \left[\lambda_{q \psi}^L P_L + \lambda_{q \psi}^R P_R\right] q \, V_\mu^\dagger - \bar{N} \gamma^\mu \left[\lambda_{q N}^L P_L + \lambda_{q N}^R P_R\right] q \, V_\mu^\dagger + \text{h.c.}
\end{equation}
The decay width for the three-body decay mediated by a heavy charged vector particle reads
\begin{equation}\label{Leffvector}
  \Gamma(\psi_\text{DM} \to q\,\bar q\, N) = 
  \frac{m_{\text{DM}}^5}{32(2 \pi)^3 m_V^4} \left\{C_1^V
  F_1(m_N^2/m_{\text{DM}}^2) + C_2^V
  F_2(m_N^2/m_{\text{DM}}^2)\right\}\;,
\end{equation}
where the functions $F_1$ and $F_2$ were defined in Eqs.~(\ref{eqn:F1}),
(\ref{eqn:F2}) and
\begin{align}
  C_1^V &\equiv \left(|\lambda_{q \psi}^L|^2 + |\lambda_{q \psi}^R|^2\right)
   \left(|\lambda_{q N}^L|^2 + |\lambda_{qN}^R|^2\right) + 
  2 \eta\,\text{Re} \left(\lambda_{q \psi}^L \lambda_{q N}^{L*} \lambda_{q \psi}^R
  \lambda_{q N}^{R*}\right),\\
  C_2^V &\equiv 2 \eta\,
  \text{Re}\left[\left(\lambda_{q \psi}^L \lambda_{q N}^{L*}\right)^2 +
  \left(\lambda_{q \psi}^R \lambda_{q N}^{R*}\right)^2\right] =
  C_2^\Sigma.
\end{align}

The energy spectrum of quarks and antiquarks produced in the three-body dark matter decay is fairly model dependent, since it depends both on the couplings which are involved in the decay as well as whether the mediating particle is a heavy charged scalar or a vector. More specifically, when the decay is mediated by a heavy charged scalar, the differential decay rate is:
\begin{eqnarray}\label{PositronSpectrumMassiveScalar}
	\frac{d\Gamma}{dE} & = & \frac{3}{2(2\pi)^3} \frac{m_{\text{DM}}E^2(E_\text{max}-E)^2}{(m_{\rm DM}-2 E)m_{\med}^4} \Bigg\{ (|\lamFL|^2+|\lamFR|^2)(|\lamNNL|^2+|\lamNNR|^2)\times  \nonumber\\
	&   & {} \frac{16E^2+2E(E_\text{max}-9m_{\rm DM})-3m_{\rm DM}(E_\text{max}-2m_{\rm DM})}{3(m_{\rm DM}-2E)^2} \\
	&   & {} + \eta \, \Re\left[\left(\lamFL \lamNNLast\right)^2 + \left(\lamFR \lamNNRast\right)^2\right] \frac{m_\nn}{m_{\rm DM}-2E} \nonumber\\
	&   & {} -2 \eta \, \Re\left(\lamFL \lamNNLast \lamFR \lamNNRast\right) \Bigg\}\;, \nonumber
\end{eqnarray}
where 
\begin{equation}
 E_\text{max} \simeq \frac{m_{\rm DM}}{2} \left(1-\frac{m_\nn^2}{m_{\rm DM}^2}\right) \;.
\end{equation}

\begin{figure}[t]
\begin{center}
\includegraphics[width=120mm]{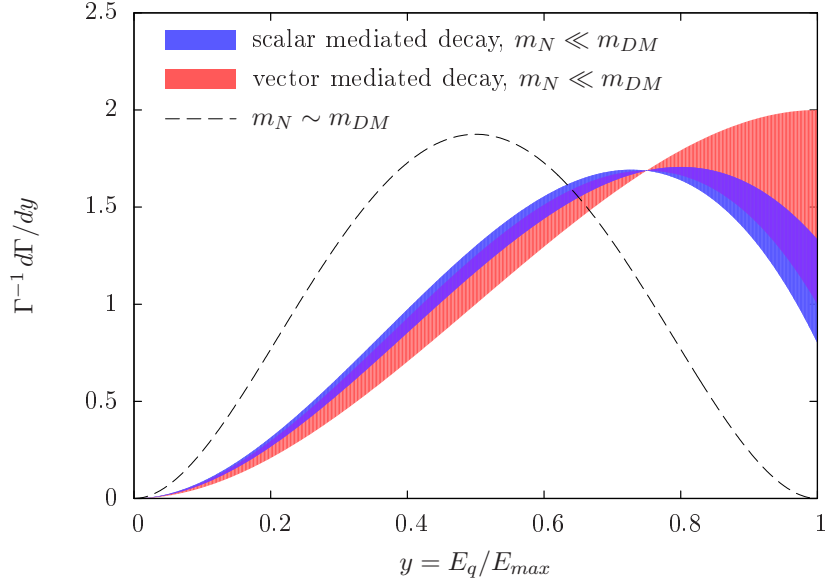}
\end{center}
\caption{\label{fig:comparison} Energy spectra of the quarks produced in the three-body decays $\psi_{\rm DM}\to q\bar q N$ mediated by a heavy scalar and a heavy vector, respectively, normalized to the total decay rate (\textit{blue:} scalar, \textit{red:} vector), when assuming that $m_N\ll m_{\rm DM}$. The black dashed line shows the spectrum in the nearly degenerate case $m_N\lesssim m_{\rm DM}$, which is identical for scalar and vector-mediated decays.}
\end{figure}

In the hierarchical limit, $m_N/m_{\text{DM}} \to 0$,
the normalized energy spectrum has a maximum and a minimum which is reached when $\lambda_{q \psi}^L=\lambda_{q \psi}^R=\lambda_{q N}^L=\lambda_{q N}^R$ and which yields
\begin{eqnarray}
\left(\frac{1}{\Gamma}\frac{d\Gamma}{dy}\right)_{\rm max,min}= 8 y^2
\times({\rm max,min})
\left\{\left(1-\frac{5}{6}y\right),
\frac{6}{5}\left(1-\frac{11}{12}y\right)\right\}\;,
\end{eqnarray}
with $y=E/E_\text{max}$. On the other hand, in the degenerate limit
$m_N/m_{\text{DM}} \to 1$, the normalized energy spectrum 
is independent of the combination of couplings and reads
\begin{eqnarray}
\left(\frac{1}{\Gamma}\frac{d\Gamma}{dy}\right)\simeq 30 y^2(1-y)^2\;.
\end{eqnarray}
The allowed ranges for the energy spectrum in the case of decays mediated by a heavy charged scalar in the hierarchical and degenerate limit are shown in Fig.~\ref{fig:comparison}.

In the case that the decay is mediated by a heavy charged vector,
the differential decay rate is:
\begin{eqnarray}\label{PositronSpectrumMassiveVector}
	\frac{d\Gamma}{dE} & = & \frac{6}{(2\pi)^3} \frac{m_{\text{DM}}E^2(E_\text{max}-E)^2}{(m_{\rm DM}-2E)m_{\medV}^4} \Bigg\{ (|\lamFL\lamNNL|^2+|\lamFR\lamNNR|^2)\times  \nonumber\\
	&   & {} \frac{16E^2+2E(E_\text{max}-9m_{\rm DM})-3m_{\rm DM}(E_\text{max}-2m_{\rm DM})}{3(m_{\rm DM}-2E)^2} \nonumber\\
	&   & {} + \eta \, \Re\left[\left(\lamFL \lamNNLast\right)^2 + \left(\lamFR \lamNNRast\right)^2\right] \frac{m_\nn}{m_{\rm DM}-2E} \nonumber\\
	&   & {} +2 \Big[ (|\lamFL\lamNNR|^2+|\lamFR\lamNNL|^2) + 2\eta\Re\left(\lamFL \lamNNLast \lamFR \lamNNRast\right)\Big] \times  \nonumber\\
	&   & {} \frac{4E^2+2E(E_\text{max}-3m_{\rm DM})-3m_{\rm DM}(E_\text{max}-m_{\rm DM})}{3(m_{\rm DM}-2E)^2}\Bigg\} \;.
\end{eqnarray}

In this case, the maximum and minimum of the energy spectrum is defined by
the two curves
\begin{eqnarray}
\left(\frac{1}{\Gamma}\frac{d\Gamma}{dy}\right)_\text{max,min}=
9y^2\times
({\rm max,min})\left\{\left(1-\frac{8}{9}y\right),\frac{2}{3}\left(1-\frac{2}{3}y\right)\right\}\;,
\end{eqnarray}
which correspond to the cases where the dark matter particle and the daughter neutral particle couple to quarks with the same chirality or with opposite chirality. On the other hand,
in the degenerate limit we find:
\begin{eqnarray}
\left(\frac{1}{\Gamma}\frac{d\Gamma}{dy}\right)\simeq 30 y^2(1-y)^2\;,
\end{eqnarray}
which is identical to the result for the decay mediated by a heavy scalar particle.

For the hierarchical limit, it is apparent from the figure that both spectra are very similar for $E\lsim 3/4 E_{\rm max}$, whereas at the endpoint of the spectrum the difference amounts at most to a factor of 2. The largest difference corresponds to the case where the dark matter particle and the heavy neutral daughter particle $N$ couple to quarks with different chiralities, namely in the scenario where $\lambda_{q \psi}^L=0$ and $\lambda_{q N}^R=0$ (or, alternatively, when $\lambda_{q \psi}^R=0$ and $\lambda_{q N}^L=0$). In contrast, it can be checked from Eqs.~(\ref{PositronSpectrumMassiveScalar}),~(\ref{PositronSpectrumMassiveVector}) that the decay spectra are identical when the dark matter particle and the heavy neutral particle $N$ couple to quarks with the same chiralities. Since the spectra are fairly similar we only analyze the case with the decay mediated by a scalar and the chiral couplings, motivated by the case when the neutral daughter particle is a neutrino.

\begin{figure}[h!]
\begin{center}
\psfrag{mdmhhhhhhhhhhh}{\scriptsize $m_\text{DM}$ [GeV]}
\psfrag{gammadmhhhh}{\scriptsize \hspace{0.2cm} $\Gamma^{-1}$ [s]}
\includegraphics[width=80mm]{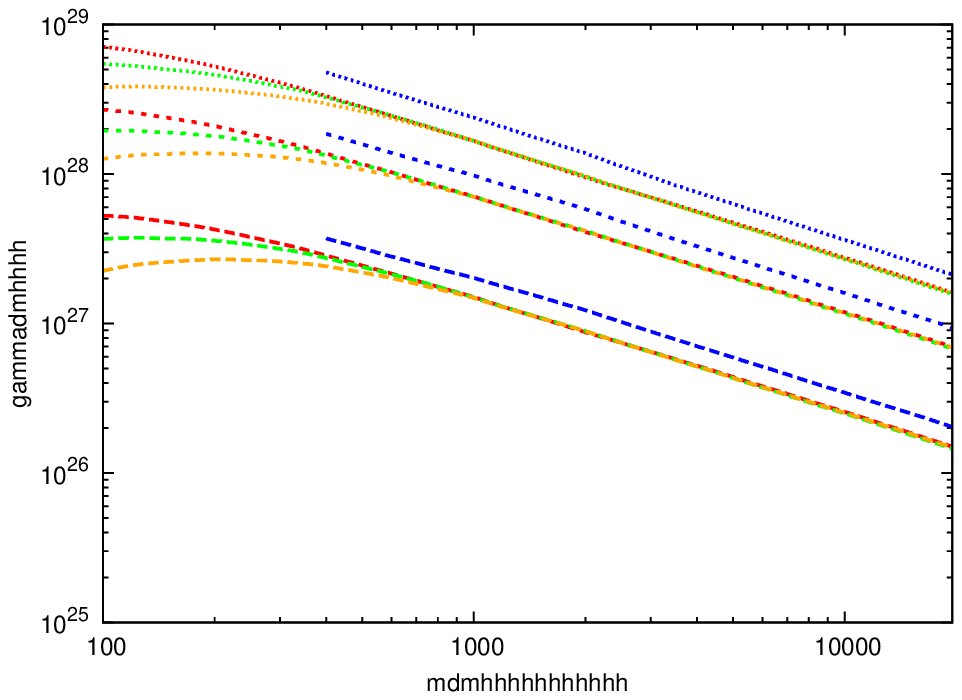}~~~\includegraphics[width=80mm]{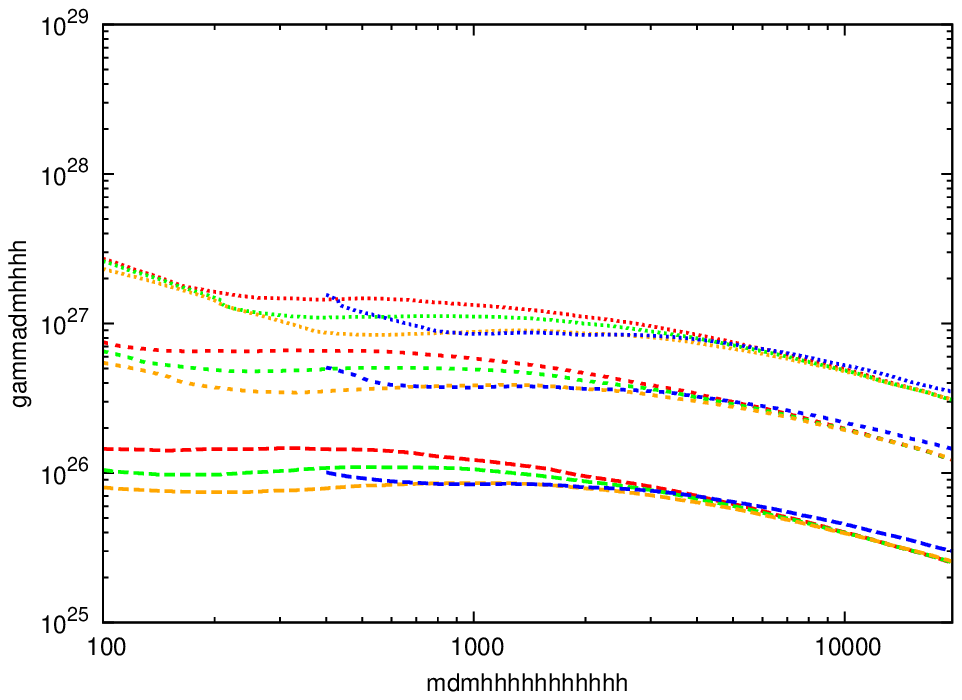}
\end{center}
\caption{\label{fig:spin12_quarks}Same as Fig.~\ref{fig:spin12_bosons}, but for the decay of a fermionic dark matter particle into a quark--antiquark pair and a neutrino. The red line corresponds to the decay $\psi_\text{DM} \to d \bar d \nu$, green to $\psi_\text{DM} \to c \bar c \nu$, orange to $\psi_\text{DM} \to b \bar b \nu$ and blue to $\psi_\text{DM} \to t \bar t \nu$.}
\end{figure}

We show in Fig.~\ref{fig:spin12_quarks} the constraints on the dark matter lifetime as a function of the mass in the case that the dark matter particle decays into a quark--antiquark pair and a neutrino. In the left-hand plot we require that the total flux, including the background antiproton flux calculated in~\cite{Bringmann:2006im}, does not exceed the antiproton-to-proton ratio observed by PAMELA. In the right-hand plot we show conservative constraints from requiring that the contribution from dark matter decay alone does not exceed the antiproton-to-proton ratio. The red line corresponds to the constraints for the decay into light quarks, concretely the decay into a down-quark pair and a neutrino, $\psi_\text{DM} \to d \bar d \nu$; the results for the decay into $u$ and $s$ quarks are very similar and are not displayed. On the other hand, the constraints from the decay into the heavy quarks $c$, $b$ and $t$ are shown, respectively, in green, orange and blue. Notice that the constraints from decays into heavy flavors are more stringent, because of the higher multiplicity of antiprotons in the fragmentation of heavy quarks compared to the decays into the lighter quark flavors. This effect is particularly important for decays into top-antitop pairs, however, this decay is only relevant for relatively large dark matter masses, $m_\text{DM} > 2 m_t \simeq 350$ GeV, for which this decay is kinematically accessible. 

\begin{figure}[h!]
\begin{center}
\psfrag{mdmhhhhhhhhhhh}{\scriptsize $p_\text{max}$ [GeV]}
\psfrag{gammadmhhhhhhhhhhhhhhhhhhhhhhhh}{\scriptsize\hspace{0.2cm}  $m_\text{DM} \, \Gamma^{-1}$ [GeV s]}
\includegraphics[width=80mm]{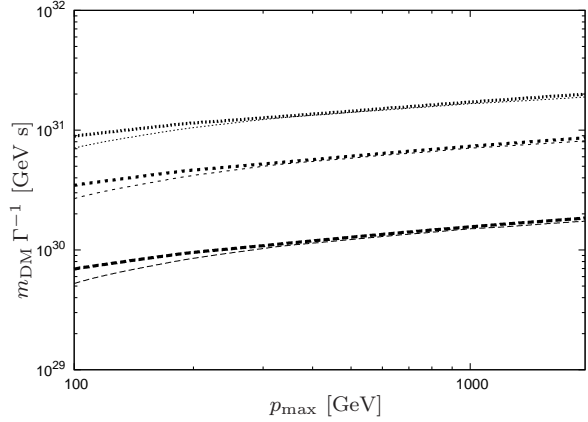}
\end{center}
\caption{\label{fig:spin12_degenerate}Direct comparison of the constraints on the nearly mass-degenerate three-body decay $\psi_\text{DM} \to d \bar{d} N$ with $m_N = 0.9 \, m_\text{DM}$ and the decay $\psi_\text{DM} \to d \bar{d} \nu$ with a massless fermion $\nu$. The thin lines correspond to the massless case, whereas the thick lines correspond to the massive case. On the  horizontal axis we plot the maximum quark momentum, while we plot the combination $m_\text{DM} \Gamma^{-1}$ on the vertical axis as explained in the text.}
\end{figure}

In Fig.~\ref{fig:spin12_degenerate} we show a comparison between the constraints resulting from the decay $\psi_\text{DM} \to d \bar{d} \nu$ with a massless fermion $\nu$ and the decay $\psi_\text{DM} \to d \bar{d} N$ with a massive fermion $N$ which is near-degenerate in mass with the dark matter particle, $m_N = 0.9 \, m_\text{DM}$. Instead of mass vs. lifetime, we show the maximum quark momentum vs lifetime times mass in this plot. Because the flux of antiprotons is proportional to $\rho_\text{DM} \Gamma/m_\text{DM}$, the combination $m_\text{DM} \Gamma^{-1}$ can be constrained independent of the dark matter mass, allowing for direct comparison between the massive and massless cases. The constraints are quite similar, although not completely identical due to the different energy spectra in the massless and massive cases.

\subsection{Decay of Spin-0 Dark Matter Particles}
\label{spin-zero-tree}

In the case of a decaying scalar dark matter particle, we consider the following decay channels:
\begin{itemize}
\item Decay into a pair of weak gauge bosons: $\phi_\text{DM} \to Z^0 Z^0$, $\phi_\text{DM} \to W^+ W^-$.
\item Decay into a pair of Higgs bosons: $\phi_\text{DM} \to h^0 h^0$.
\item Decay into a quark--antiquark pair: $\phi_\text{DM} \to q_i\bar q_j$.
\end{itemize}
To calculate the antiproton fluxes at Earth we again follow the procedure outlined in subsection \ref{spin-half-tree}.

\begin{figure}[h!]
\begin{center}
\psfrag{mdmhhhhhhhhhhh}{\scriptsize $m_\text{DM}$ [GeV]}
\psfrag{gammadmhhhh}{\scriptsize\hspace{0.2cm}  $\Gamma^{-1}$ [s]}
\includegraphics[width=80mm]{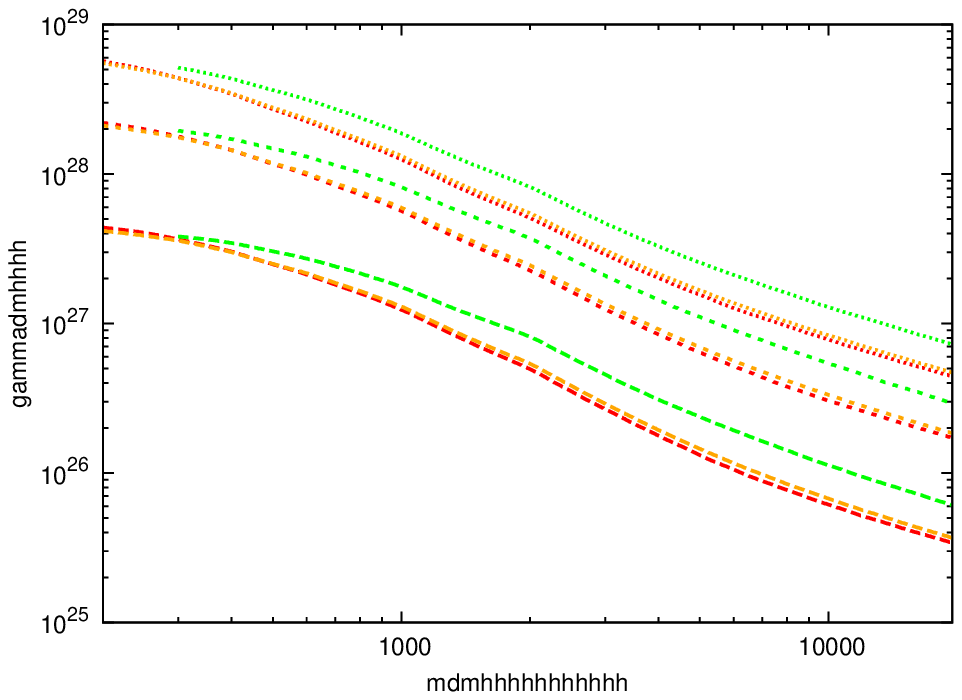}~~~\includegraphics[width=80mm]{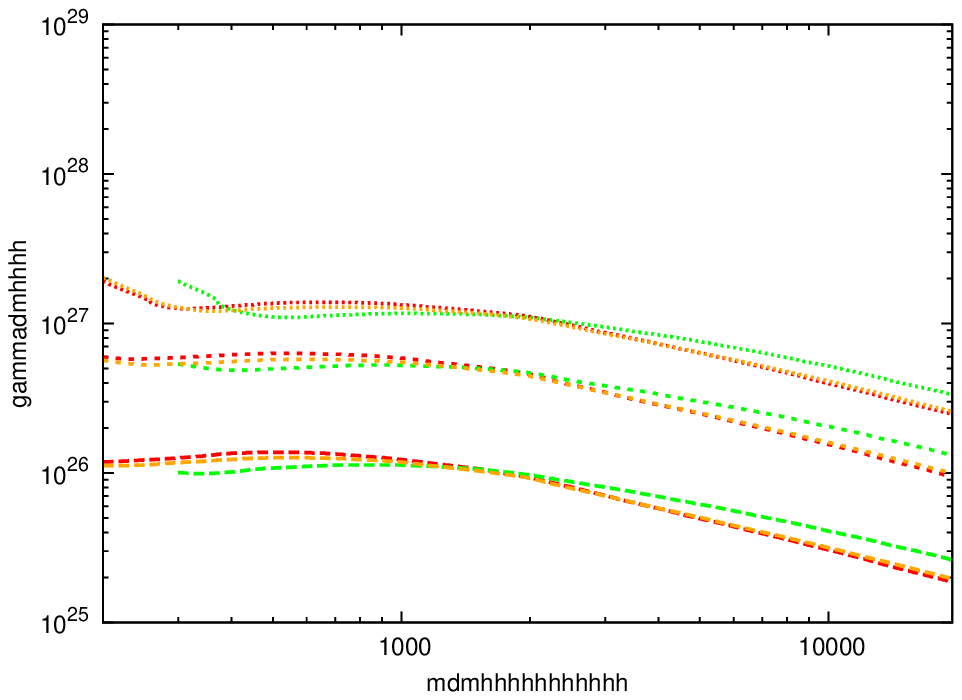}
\end{center}
\caption{Lower limits on the partial  lifetime of a scalar dark matter particle which decays into two bosons from the requirement of not exceeding the antiproton-to-proton fraction as measured by PAMELA at 95\% C.L. The red line corresponds to the decay  $\phi_\text{DM} \to W^\pm W^\mp$, the orange line to $\phi_\text{DM} \to Z^0 Z^0$  and the green line to $\phi_\text{DM} \to h^0 h^0$, while the long-dashed, short-dashed and dotted lines to the MIN, MED and MAX propagation models, respectively (see text for details). \textit{Left panel:} Constraints for primary antiprotons from dark matter decay + secondary antiprotons from cosmic-ray spallation. \textit{Right panel:} Constraints for primary antiprotons from dark matter decay only.}
\label{fig:spin0_bosons}
\end{figure}

We show in Fig.~\ref{fig:spin0_bosons} the resulting constraints for the scalar dark matter decay modes $\phi_\text{DM} \to W^\pm W^\mp$, $\phi_\text{DM} \to Z^0 Z^0$ and $\phi_\text{DM} \to h^0 h^0$, as red, orange and green lines respectively, where we again assume a Standard Model Higgs boson with a mass $m_{h^0} = 125 \GeV$.

\begin{figure}[h!]
\begin{center}
\psfrag{mdmhhhhhhhhhhh}{\scriptsize $m_\text{DM}$ [GeV]}
\psfrag{gammadmhhhh}{\scriptsize \hspace{0.2cm} $\Gamma^{-1}$ [s]}
\includegraphics[width=80mm]{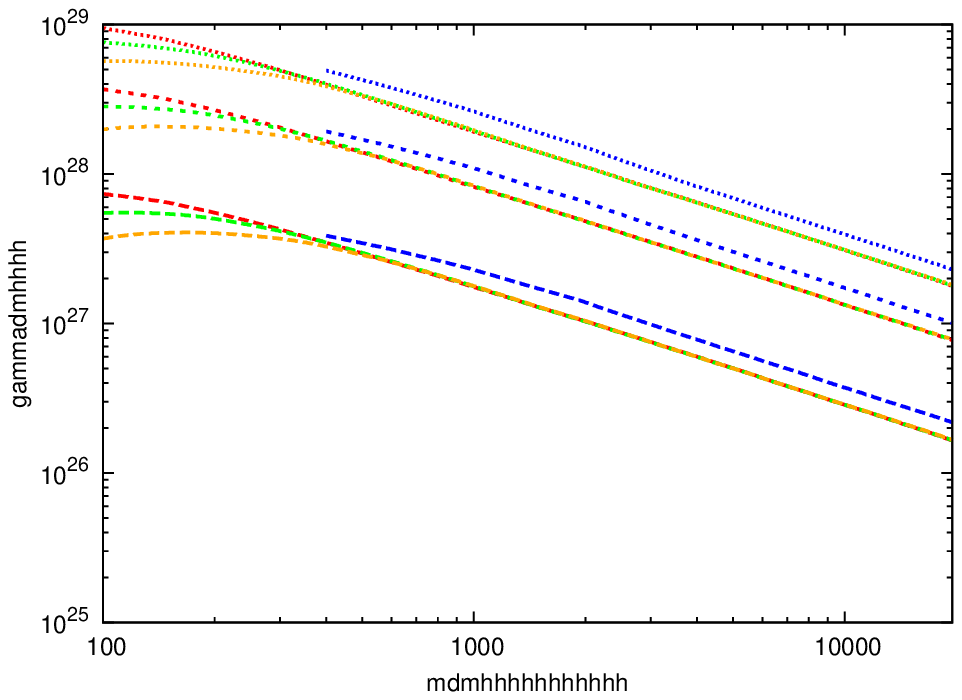}~~~\includegraphics[width=80mm]{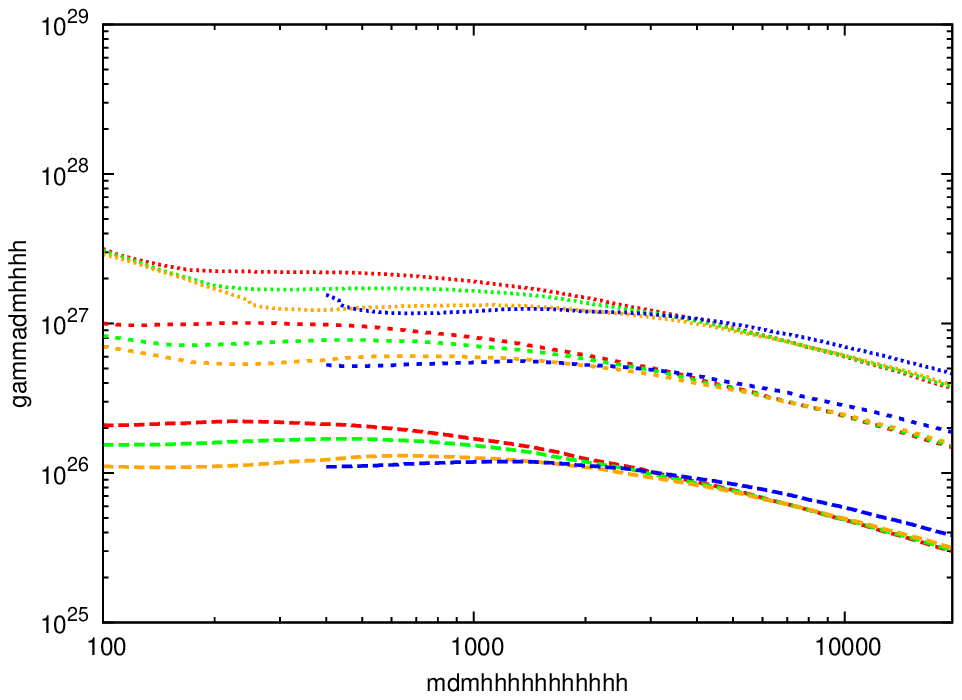}
\end{center}
\caption{Same as Fig.~\ref{fig:spin0_bosons}, but for the decay of a scalar dark matter particle into a quark--antiquark pair. The red line corresponds to the decay  $\phi_\text{DM} \to d \bar{d}$, green to $\phi_\text{DM} \to c \bar{c}$, orange to $\phi_\text{DM} \to b\bar{b}$ and blue to $\phi_\text{DM} \to t \bar{t}$.}
\label{fig:spin0_quarks}
\end{figure}

The constraints for decays into quark--antiquark pairs, on the other hand, are shown in Fig.~\ref{fig:spin0_quarks}. We show in red the constraints on the lifetime for the decay mode $\phi_\text{DM} \to d \bar{d}$, in green for $\phi_\text{DM} \to c \bar{c}$, in orange for $\phi_\text{DM} \to b\bar{b}$ and in blue for $\phi_\text{DM} \to t \bar{t}$; the limits for decays into the light quarks $u$ and $s$ are very similar to the limits for decays into $d$ quarks and are not displayed.

\section{Constraints from Radiative Decays into Gamma Rays}
\label{sec:gamma_rays}
A dark matter particle that decays into electrically charged particles at tree-level
generically also possesses a decay mode producing monoenergetic photons that is
induced by radiative corrections. As argued in \cite{Garny:2010eg}, although
the radiative decay mode is loop-suppressed, gamma-ray line searches
can give rise to relevant constraints on the dark matter lifetime for leptonically
decaying dark matter. Here, we extend this investigation to the case of hadronically
decaying dark matter. More specifically, we analyze the constraints on the parameter
space of decaying dark matter scenarios where the dark
matter particle has spin 1/2 and decays into a quark--antiquark pair
and a neutral daughter particle at tree level. In that case, a two-body decay producing
a monoenergetic photon is induced at the one-loop level. The resulting constraint on the
lifetime will thus be {\it independent} of the details of the
propagation of the antiprotons from their production point to
the Earth, which is the most important source of uncertainty
in the constraints derived in the previous section. The constraints,
however, will depend on the details of the underlying particle
physics model, namely on whether the decay is mediated by a heavy
charged scalar or by a heavy charged vector particle, as we will show below.

We first consider the three-body decay of spin-1/2 dark matter particles into a quark--antiquark pair and a neutral daughter particle, mediated by the exchange of a heavy charged scalar particle. The part of the Lagrangian which induces such decays is given in Eq.~(\ref{Leffscalar}). From this Lagrangian, the decay rate of the loop-induced process $\psi_\text{DM} \to \gamma N$ can be calculated in a straightforward manner. In the limit
$m_N,m_q\ll m_{\text{DM}} \ll m_\Sigma$ the result reads
\begin{align} \label{eq:radiative-scalar}
  \Gamma(\psi_\text{DM} \to \gamma N) = {}& \frac{ e^2}{8\pi
  \left(64\pi^2\right)^2} \frac{m_{\text{DM}}^5}{m_\Sigma^4} \left(1 -
  \frac{m_N^2}{m_{\text{DM}}^2}\right)^3 \left(1 - \frac{\eta \,
  m_N}{m_{\text{DM}}}\right)^2 \nonumber\\ & \times \left[\sum_q Q_q
  \left(\lambda_{q N}^L \lambda_{q \psi}^L - \eta \, \lambda_{q N}^R
  \lambda_{q \psi}^R\right)\right]^2\;,
\end{align}
where $q=\{u,d,c,s,b,t\}$ runs over all quark flavors which interact with the dark matter particle and the neutral daughter particle, and $Q_q$ is the charge of the quark in the loop in units of the electron charge ($Q_q=2/3$ for up-type quarks and $Q_q=-1/3$ for down-type quarks).

Likewise, when the decay is mediated by a heavy charged vector, corresponding to the Lagrangian in Eq. (\ref{Leffvector}), the decay rate of the loop-induced process $\psi_\text{DM} \to \gamma N$ reads
\begin{align}  \label{eq:radiative-vector}
  \Gamma(\psi_\text{DM} \to \gamma N) &= \frac{9 e^2}{8\pi
  \left(32\pi^2\right)^2} \frac{m_{\text{DM}}^5}{m_V^4} \left(1 -
  \frac{m_N^2}{m_{\text{DM}}^2}\right)^3 \left(1 - \frac{\eta \,
  m_N}{m_{\text{DM}}}\right)^2 \nonumber\\ & ~~~ \times \left[\sum_q
  Q_q \left(\lambda_{q N}^L \lambda_{q \psi}^L - \eta \, \lambda_{q N}^R
  \lambda_{q \psi}^R\right)\right]^2\;,
\end{align}
where, as before, we have assumed the mass hierarchy $m_N,m_q\ll m_{\text{DM}} \ll m_\Sigma$.

\subsection{Neutrinos as Neutral Daughter Particles}

Let us analyze first the case where the neutral daughter particle is a left-handed neutrino, implying $\lambda^R_{qN} = 0$ and $m_N \simeq 0$. It then follows from Eq.~(\ref{eq:radiative-scalar}) that in order to obtain a sizeable decay rate into monoenergetic gamma rays, it is necessary that the dark matter particles also couple to quarks with left-handed chirality. Otherwise, the decay rate is proportional to the quark mass squared, which greatly reduces the intensity of the gamma-ray line unless the quark circulating in the loop is a top quark. If $\lambda^L_{q \psi}\neq 0$, we find
\begin{align}
{\rm BR}(\psi_\text{DM} \to \gamma \nu)&\simeq  
\frac{9\alpha_\text{em}}{8\pi}
\frac{|\sum_q Q_q \lambda_{q N}^L \lambda_{q \psi}^L|^2}{\sum_q |\lambda_{q N}^L \lambda_{q \psi}^L|^2}&\text{for intermediate charged scalar,}\nonumber \\ 
{\rm BR}(\psi_\text{DM} \to \gamma \nu)&\simeq  
\frac{81\alpha_\text{em}}{8\pi}
\frac{|\sum_q Q_q \lambda_{q N}^L \lambda_{q \psi}^L|^2}{\sum_q |\lambda_{q N}^L \lambda_{q \psi}^L|^2} &\text{for intermediate charged vector.}
\end{align}
We list in Table \ref{tab:BRs} the branching ratios for a series of interesting scenarios where the dark matter particle decays into just a single flavor of down-type or up-type quarks ($\psi_\text{DM} \to d\bar d \nu$ and $\psi_\text{DM} \to u\bar u \nu$ respectively), democratically into all three quarks of the same type, either down type or up-type ($\psi_\text{DM} \to q_d\bar q_d \nu$ and $\psi_\text{DM} \to q_u\bar q_u \nu$, respectively), or democratically into all quark flavors ($\psi_\text{DM} \to q \bar q \nu$). It is apparent from the table that in this scenario the relative branching ratio ranges between $\alpha_\text{em}/(8\pi)\simeq 3\times 10^{-4}$ and $108 \, \alpha_\text{em}/(8\pi)\simeq 0.03$, where decays mediated by virtual vector particles generally have a larger ratio of two-body decays to three-body decays than those mediated by virtual scalars. 
The PAMELA measurements on the antiproton to proton fraction constrains the inverse decay width of the three-body decays to be smaller than $\sim 10^{27}$ s, as shown in the previous section. Therefore, the inverse decay rate of the radiatively induced process $\psi\to \gamma N$ will be in the range $10^{29}-10^{31}$ s. The present observational upper bound on the inverse decay width into monoenergetic gamma rays is about $5\times 10^{28} \dots 10^{29}~{\rm s}$ for dark matter masses in the range $2-600$ GeV. Furthermore, in the future the projected Cherenkov Telescope Array (CTA)~\cite{Consortium:2010bc} will constrain the inverse decay width in the TeV mass range also at the level $5\times 10^{28} \dots 10^{29}~{\rm s}$ from observations of the diffuse electron background, assuming an observational time of $T=1000~{\rm h}$ and an energy resolution of 10\%~\cite{Garny:2010eg}. Therefore, for some concrete particle physics scenarios, the present and projected searches of the radiatively induced gamma-ray lines could provide constraints on those scenarios which are competitive with the ones stemming from the tree-level antiproton production.

\begin{table}
\begin {center}
  \begin{tabular}{c|c|c}
  Channel & intermediate scalar & intermediate vector \\
  \hline
  $\psi\to e^+e^-\nu$ & $3\alpha_\text{em}/(8\pi)$ & $27\alpha_\text{em}/(8\pi)$ \\
  $\psi\to \ell^+\ell^-\nu$ & $9\alpha_\text{em}/(8\pi)$ & $81\alpha_\text{em}/(8\pi)$ \\
  $\psi\to d\bar d\nu$ & $\alpha_\text{em}/(8\pi)$ & $9\alpha_\text{em}/(8\pi)$ \\
  $\psi\to u\bar u\nu$ & $4\alpha_\text{em}/(8\pi)$ & $36\alpha_\text{em}/(8\pi)$ \\
  $\psi\to q_d\bar q_d\nu$ & $3\alpha_\text{em}/(8\pi)$ & $27\alpha_\text{em}/(8\pi)$ \\
  $\psi\to q_u\bar q_u\nu$ & $12\alpha_\text{em}/(8\pi)$ & $108\alpha_\text{em}/(8\pi)$ \\
  $\psi\to q\bar q\nu$ & $3\alpha_\text{em}/(16\pi)$ & $27\alpha_\text{em}/(16\pi)$
  \end{tabular}
  \end{center}
  \caption{Ratio $\Gamma(\psi\to\gamma\nu)/\sum_f\Gamma(\psi\to f\bar f\nu)$ of the one-loop decay rate into monoenergetic photons to the tree-level decay rate into fermions, for the case of scalar and vector mediated
  decays (neglecting fermion masses, and assuming purely chiral couplings). The decay channel $\psi\to \ell^+\ell^-\nu$ corresponds to a dark matter particle which decays into $e^+e^-\nu$, $\mu^+\mu^-\nu$ and $\tau^+\tau^- \nu$ with equal rates (flavor-democratic decay). Similarly, the last three lines correspond to flavor-democratic decay into the down-type quarks, the up-type quarks, or all quarks, respectively (assuming
  that $m_\text{DM}\gg 2m_t$ in the latter two cases). The ratios are in the permille or percent range ($\alpha_\text{em}=e^2/(4\pi) \approx 1/137$).}
  \label{tab:BRs}
\end{table}

\begin{figure}
\begin{center}
\psfrag{mdmhhhhhhhhhhh}{\scriptsize $m_\text{DM}$ [GeV]}
\psfrag{gammadmhhhhhhhhhhhhhhhhhhhhhhhh}{\scriptsize \hspace{.2cm} $\Gamma^{-1}(\psi_\text{DM} \to d \bar{d} \nu)$ [s]}
\includegraphics[width=80mm]{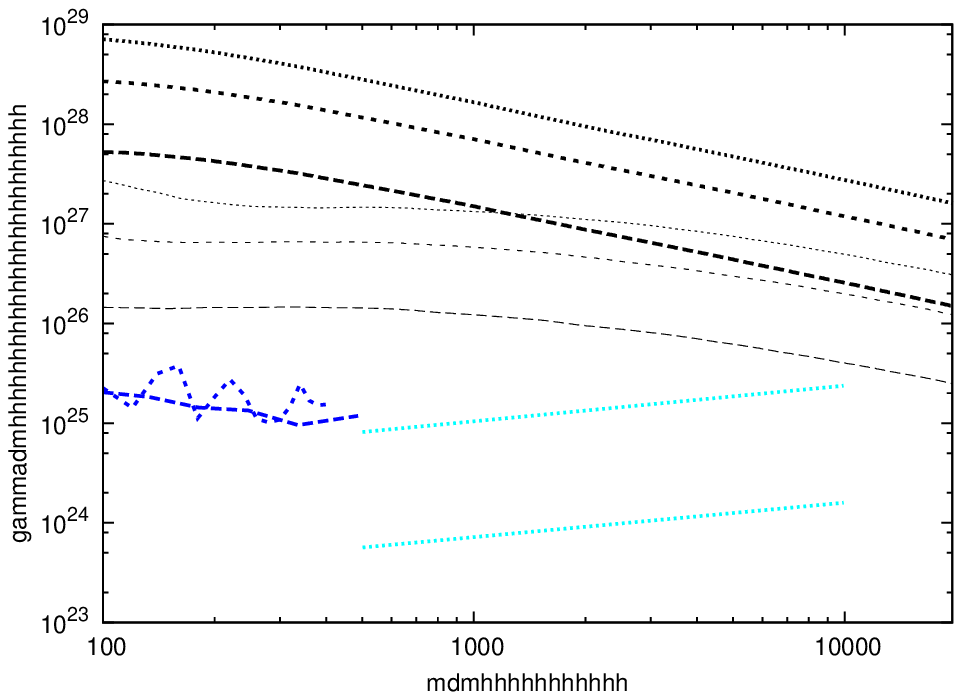}~~~\includegraphics[width=80mm]{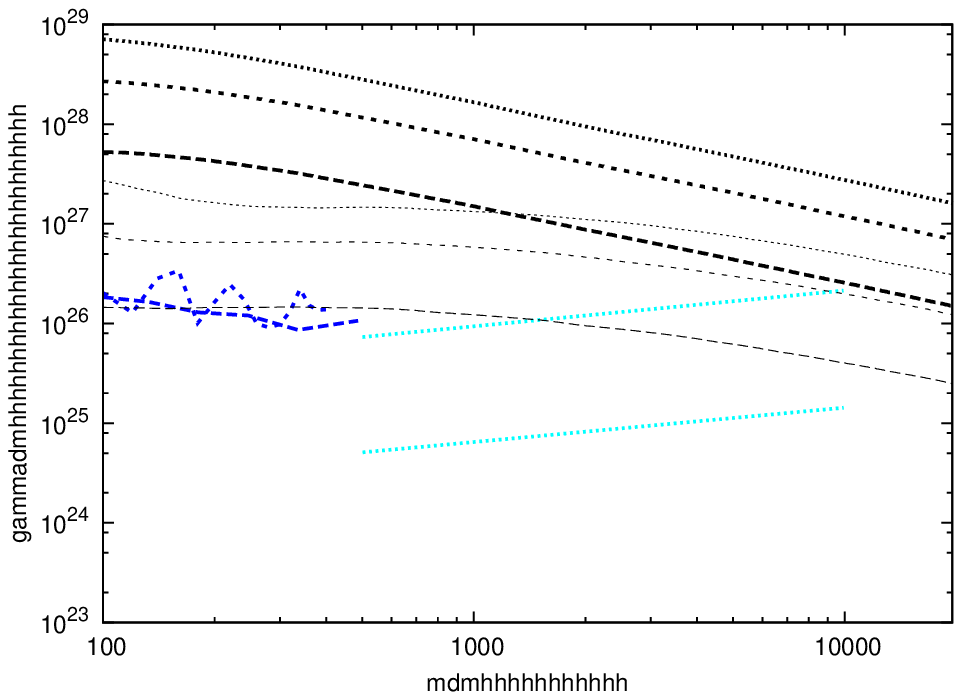} \\
\psfrag{gammadmhhhhhhhhhhhhhhhhhhhhhhhh}{\scriptsize  \hspace{.2cm} $\Gamma^{-1}(\psi_\text{DM} \to q_u \bar{q_u} \nu)$ [s]}
\includegraphics[width=80mm]{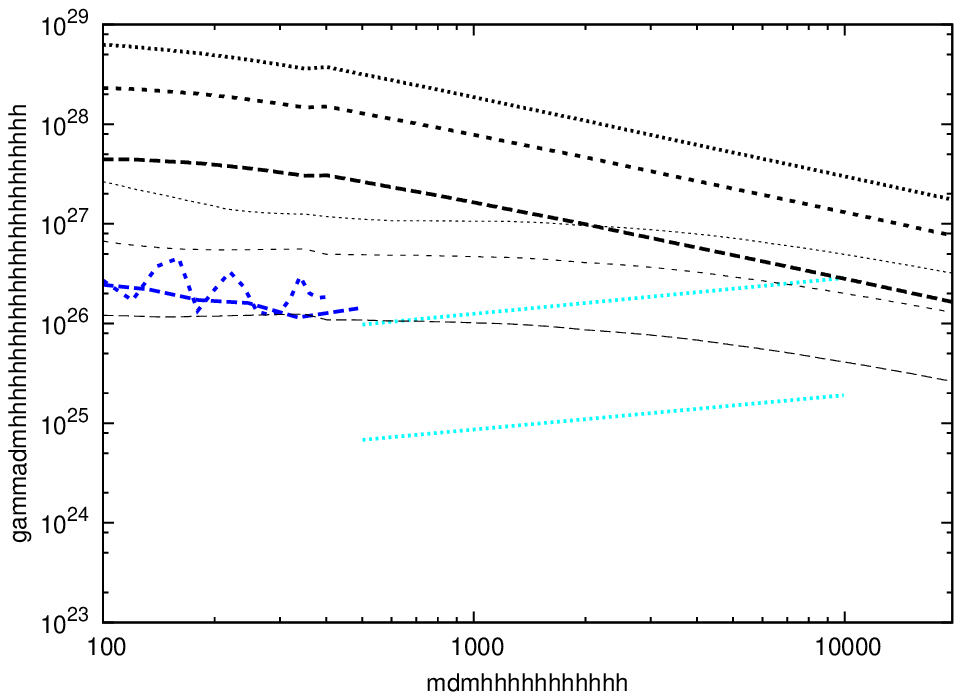}~~~\includegraphics[width=80mm]{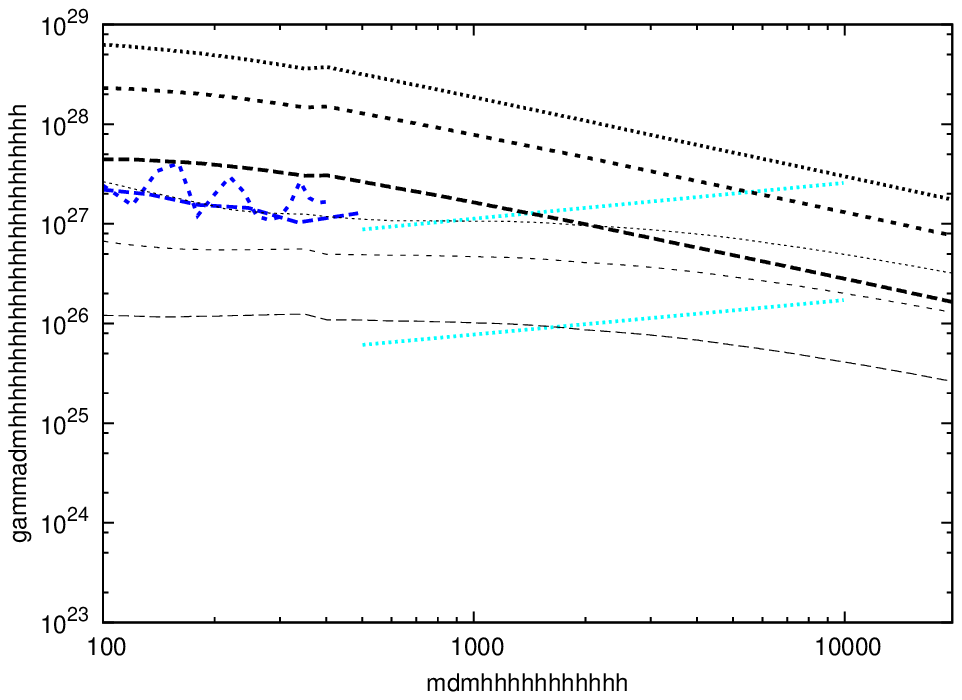} 
\end{center}
\caption{Comparison of constraints from cosmic-ray antiprotons and radiatively generated gamma-ray lines for the tree-level decay into a pair of down-type quarks and a neutrino ($\psi_\text{DM} \to d \bar d \nu$, top panel), as well as for the democratic decay into charge $2/3$ quark flavors ($\psi_\text{DM} \to q_u \bar{q}_u \nu$, bottom panel). The thick black lines are the limits on the decay width from antiprotons including the background from cosmic-ray spallation, while the thin lines correspond to the limits resulting from dark matter decay only. On the other hand, the blue lines represent the limits from gamma-ray lines from the Fermi LAT data~\cite{Abdo:2010nc,Vertongen:2011mu}, while the cyan lines represent the projected limits for the Cherenkov Telescope Array~\cite{Garny:2010eg}. \textit{Left panels:} Intermediate scalar. \textit{Right panels:} Intermediate vector.}
\label{fig:line}
\end{figure}

The interplay between gamma-ray constraints and antiproton constraints is illustrated in Fig.~\ref{fig:line} for the scenarios where the dark matter particle decays into down quarks and a neutrino ($\psi_\text{DM} \to d \bar{d} \nu$, top panel)  or democratically into all three charge $2/3$ quark flavors ($\psi_\text{DM} \to q_u \bar{q}_u \nu$, bottom panel), and which yield, respectively, the minimum and maximum ratio for the radiative two-body decays to the tree-level three-body decays  ({\it cf.} Tab.~\ref{tab:BRs}). In the plots, we show the constraints obtained in Section~\ref{sec:tree_level} together with the constraints on gamma-ray lines from two different analyses of the Fermi LAT data (dotted blue~\cite{Abdo:2010nc}, solid blue~\cite{Vertongen:2011mu}) as well as the prospected limits at the Cherenkov Telescope Array (cyan)~\cite{Garny:2010eg}. In the plots, we use the upper limits on the inverse decay rate into monochromatic photons, and we scale them by the appropriate ratios for the radiative two-body decays to the tree-level three-body decays listed in Tab.~\ref{tab:BRs}.

The limits published recently by the Fermi LAT collaboration \cite{Ackermann:2012qk} are very similar to the ones shown  in Fig.~\ref{fig:line} within the energy range of interest to us. Whether the feature in the gamma-ray spectrum near the Galactic center at $E_\gamma\approx 130$ GeV reported in \cite{Weniger:2012tx,Bringmann:2012vr} can be interpreted in terms of decaying dark matter remains to be seen \cite{Profumo:2012tr,Boyarsky:2012ca}, and we will not pursue this issue in further detail here. We just remark that the limits arising from hadronic decay channels discussed here constitute an important consistency check for any model aiming to explain the gamma-ray feature.

\subsection{Degenerate Scenario}

In concrete particle physics models, the neutral daughter particle produced in the three-body decay might not be massless, but could instead have a sizeable mass, possibly comparable to the dark matter mass. If this is the case, the branching ratio of the radiative decay into photons could be significantly enhanced by kinematic effects, as discussed in \cite{Garny:2010eg}.

In the case of chiral couplings, for instance when the dark matter particle and the neutral daughter
particle both couple to quarks with left-handed chirality, the branching ratio into monoenergetic
gamma rays reads
\begin{align}
{\rm BR}(\psi_\text{DM} \to \gamma \nn)\simeq {} &
\frac{ 45  \alpha_\text{em}   }{8 \pi} 
\left(1-\frac{m_N^2}{m_{\text{DM}}^2}\right)^{-2} \left(1 - \frac{\eta \,  m_N}{m_{\text{DM}}}\right)^2 \nonumber\\
& \times \frac{\left[\sum_q Q_q  \left(\lambda_{q N}^L \lambda_{q \psi}^L\right)\right]^2}
     {\sum_q \left\{ 2|\lambda_{q \psi}^L|^2 |\lambda_{q N}^L|^2+\eta\,
      \text{Re}\left[\left(\lambda_{q \psi}^L \lambda_{q N}^{L*}\right)^2 \right]\right\}} \,,
\end{align}
when the decay is mediated by a heavy charged scalar, and
\begin{align}
{\rm BR}(\psi_\text{DM} \to \gamma \nn)\simeq  
{} & \frac{ 405  \alpha_\text{em}   }{8 \pi} 
\left(1-\frac{m_N^2}{m_{\text{DM}}^2}\right)^{-2} \left(1 - \frac{\eta \,  m_N}{m_{\text{DM}}}\right)^2 \nonumber\\
& {} \times \frac{\left[\sum_q Q_q  \left(\lambda_{q N}^L \lambda_{q \psi}^L\right)\right]^2}
     {\sum_q \left\{ 2|\lambda_{q \psi}^L|^2 |\lambda_{q N}^L|^2+\eta\,
      \text{Re}\left[\left(\lambda_{q \psi}^L \lambda_{q N}^{L*}\right)^2 \right]\right\}} \,,
\end{align}
when the decay is mediated by a heavy charged vector.\footnote{Note that we have used the asymptotic
approximation Eq.~(\ref{Fs-degenerate}) for the kinematical functions $F_i(x)$ for
obtaining these expressions, i.e. they are valid for $m_\nn \simeq m_{\rm DM}$. We have checked that, using the full $F_i(x)$, the resulting branching ratio smoothly interpolates between the expressions given here  for $m_\nn \simeq m_{\rm DM}$ and those for $m_\nn\to 0$ derived in the previous subsection.}
In both cases, one finds a significant
enhancement of the decay width into monoenergetic gamma rays as $m_N\to m_{\rm DM}$
when the dark matter particle and the daughter particle have opposite \textsl{CP} parities, $\eta=-1$.
In this case the branching ratio has the proportionality
\begin{equation}
{\rm BR}(\psi_\text{DM} \to \gamma \nu)\propto \left(1-\frac{m_N}{m_{\text{DM}}}\right)^{-2} \;,
\end{equation}
which can lead to a significant enhancement of the radiative decay as $m_N \to m_{\text{DM}}$.

\begin{figure}
\begin{center}
\psfrag{mdmhhhhhhhhhhh}{\scriptsize $m_\text{DM}$ [GeV]}
\psfrag{gammadmhhhhhhhhhhhhhhhhhhhhhhhh}{\scriptsize  \hspace{.2cm} $\Gamma^{-1}(\psi_\text{DM} \to d \bar{d} N)$ [s]}
\includegraphics[width=80mm]{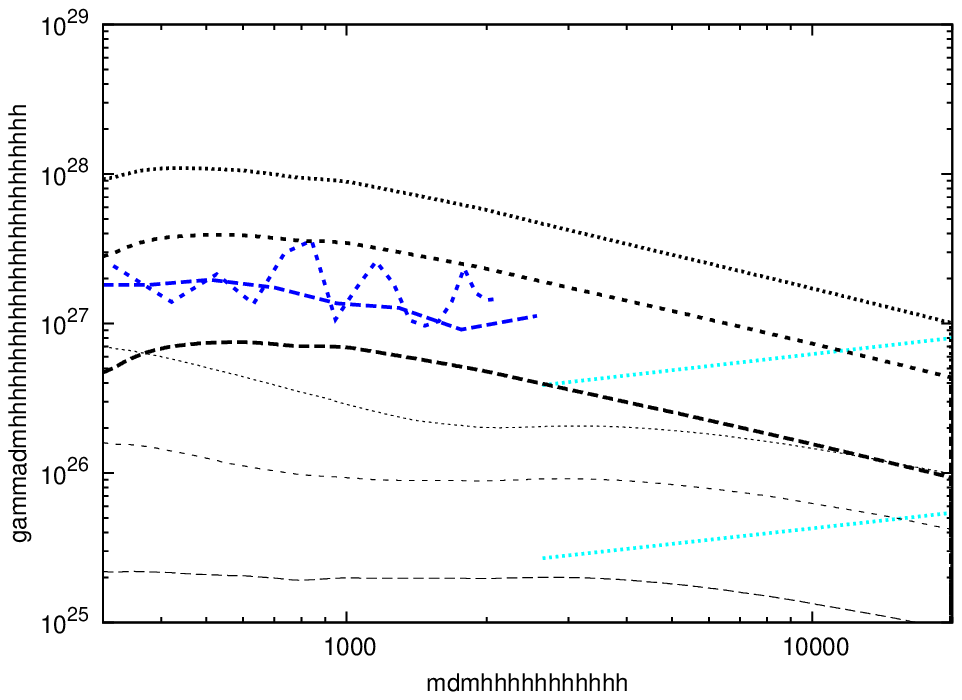}~~~\includegraphics[width=80mm]{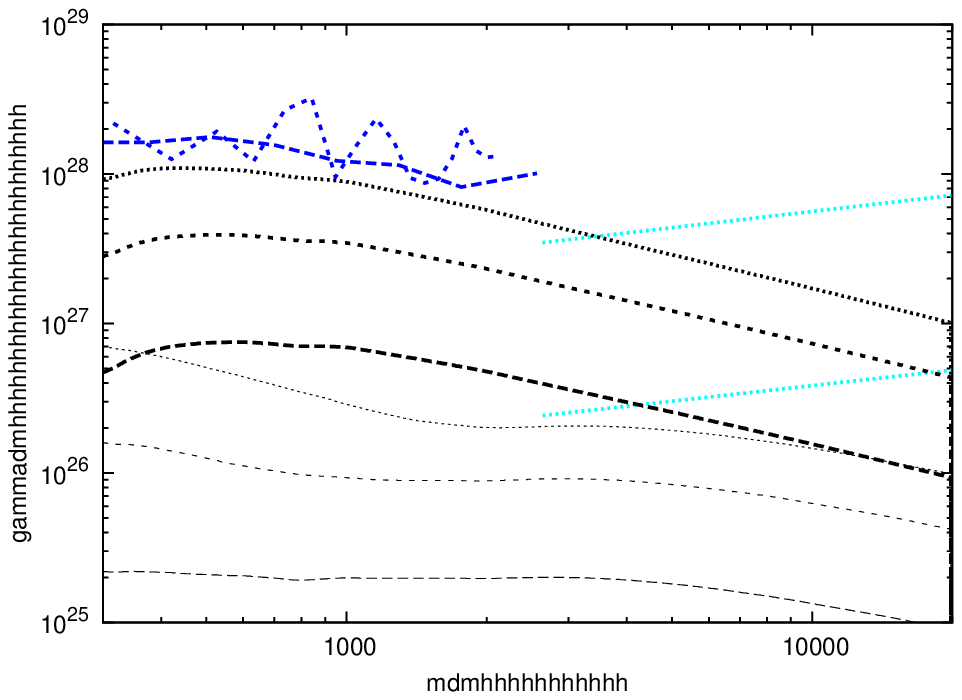}
\end{center}
\caption{Same as Fig.~\ref{fig:line}, top panels, but for the near-degenerate case $m_N = 0.9 \, m_{\text{DM}}$. Comparison of constraints from cosmic-ray antiprotons and radiatively generated gamma-ray lines for the tree-level decay into down-type quarks ($\psi_\text{DM} \to d \bar d N$) in the degenerate case where $m_N = 0.9 \, m_{\text{DM}}$. The corresponding gamma-ray constraints are stretched in the $m_\text{DM}$-direction because the available gamma-ray energy is lower if $N$ is massive. Furthermore, the constraints are scaled in the $\Gamma^{-1}$ direction due to the kinematic enhancement resulting from the opposite \textsl{CP} parities of $\psi_\text{DM}$ and $N$, as discussed in the text.}
\label{fig:line_degenerate}
\end{figure}

In Fig.~\ref{fig:line_degenerate} we show the constraints from gamma-ray and charged cosmic-ray observations for the case that the dark matter particle $\psi_\text{DM}$ and the neutral fermion $N$ are nearly mass-degenerate, $m_N = 0.9 \, m_{\text{DM}}$, and have opposite \textsl{CP} parities, $\eta=-1$. Due to the enhancement of the radiative decay mode in this case, the lower limits coming from the loop-induced decay into monochromatic photons are comparable to the lower limits obtained from the primary antiproton flux produced in the tree-level decays for scalar-mediated decays and for MIN or MED propagation parameters. For vector-mediated decays, the gamma line limits are even stronger than the antiproton limits obtained for the MAX propagation parameters.

\section{Constraints from Radiative Decays into Weak Gauge Bosons}
\label{sec:weak_gauge_bosons}

It has been proposed that the observation of cosmic ray anomalies in positron and electron fluxes reported by the PAMELA and Fermi satellite experiments could be due to dark matter that decays or annihilates primarily into leptonic channels \cite{Adriani:2008zr,Abdo:2009zk} (see, \textit{e.g.}, Ref.~\cite{Cirelli:2012ut} for a recent analysis of this scenario). Nevertheless, even leptophilic dark matter will lead to a certain amount of antiproton production induced by electroweak bremsstrahlung, as has been argued, \textit{e.g.}, in Refs.~\cite{Kachelriess:2009zy,Ciafaloni:2010ti}. We consider an
additional source of antiproton production resulting from radiative decays into weak
gauge bosons. We first consider the case of a fermionic dark matter particle and then the case of scalar dark matter.

\subsection{The Decay $\psi_\text{DM} \to Z^0 N$}

We consider a fermionic dark matter particle that decays into a pair of leptons and a
neutral particle, $\psi_\text{DM}\to\ell^+\ell^- N$, at tree level. At one-loop, the decay $\psi_\text{DM} \to \gamma N$ is
induced, allowing for constraints from gamma-ray searches~\cite{Garny:2010eg}. In addition, the decay $\psi_\text{DM} \to Z^0 N$ into a $Z$ boson is generically induced, which gives rise to antiproton production. Depending on the model,
the decay mode $\psi_\text{DM} \to W^\pm \ell^\mp$ can also be induced.
In order to study the radiative decays into weak gauge bosons, it is necessary to
specify the $SU(2)_L$-structure of the model. Here we investigate a scenario  where the
dark matter particle $\psi_\text{DM}$ and its neutral decay product $N$ are both Standard Model singlets,
and consider the effective interaction Lagrangian
\begin{eqnarray}\label{LeffSingletDM}
   \mathcal{L}_\text{eff} & = & - \lambda_{\ell \psi}^L \bar\psi_\text{DM} L \Sigma^\dag - \lambda_{\ell \psi}^R \bar\psi_\text{DM} \ell_R \sigma^\dag 
                       - \lambda_{\ell N}^L \bar\nn L \Sigma^\dag - \lambda_{\ell N}^R \bar\nn \ell_R \sigma^\dag
                       + \mbox{h.c.} \; ,
\end{eqnarray}
where $L=\begin{pmatrix} \nu_\ell ~ \ell_L\end{pmatrix}$ is the left-handed lepton doublet and $\ell_R$ is the right-handed charged lepton field. The heavy scalar fields $\Sigma$ and $\sigma$
mediate the decay, and have the same quantum numbers as $L$ and $\ell_R$, respectively. In a supersymmetric
context, they can be identified with the left- and right-handed slepton fields. The effective Lagrangian leads
to an interaction of the form described by Eq.~(\ref{Leffscalar}), except that two mediating fields, namely the charged
component of $\Sigma$ as well as $\sigma$, contribute. In addition a similar interaction involving neutrinos
and the neutral component of $\Sigma$ is present. For simplicity, we
neglect mixing of the mediating fields, i.e. we assume that $\Sigma$ and $\sigma$
coincide with mass eigenstates, and we also assume that the vacuum expectation value $\langle\Sigma^0\rangle=0$.

\begin{figure}[h!]
  \begin{center}
    \includegraphics{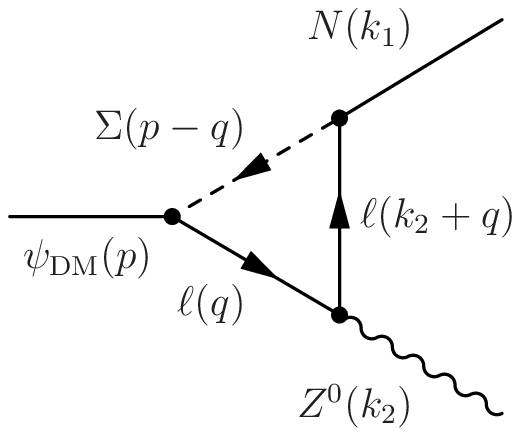}
    ~~~~~
    \includegraphics{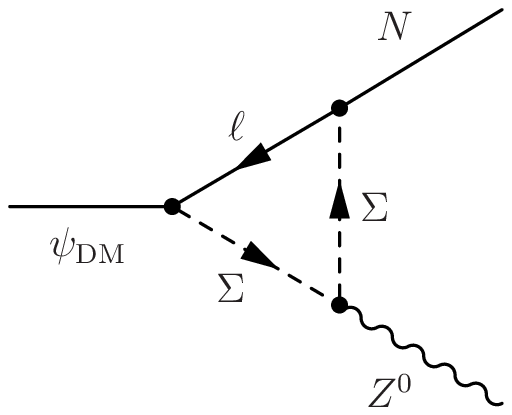}\\

    \

    \includegraphics{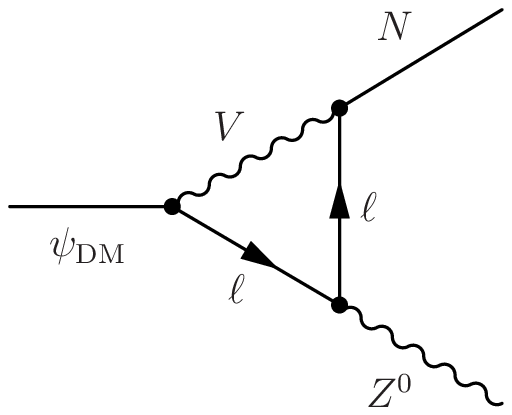}
    ~~~~~
    \includegraphics{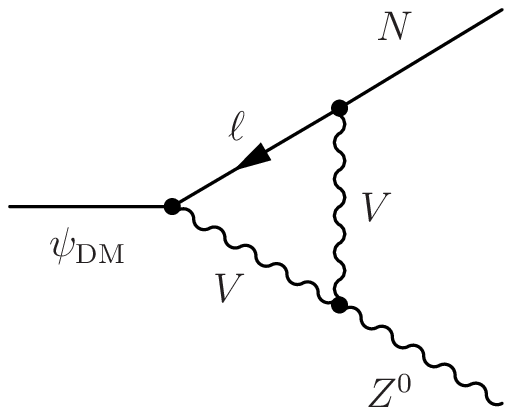}\\
    \caption{Diagrams contributing at one loop to the radiative two-body decay
    $\psi_\text{DM} \rightarrow Z^0 N$, induced by a charged scalar
    $\Sigma$ (top row) and a vector particle $V$ (bottom row), respectively.
    There are two additional diagrams in each case which differ only by the
    direction of the charge flow.}
    \label{fermion_loop}
  \end{center}
\end{figure}

The decay rates for the tree-level decays $\psi_\text{DM}\to \ell^+\ell^- N$ and $\psi_\text{DM}\to \nu\bar\nu N$ as
well as the loop-induced decay $\psi_\text{DM}\to\gamma N$ can be adapted from Ref.~\cite{Garny:2010eg}.
In addition, in this scenario the decay $\psi_\text{DM}\to Z^0 N$ occurs at the one loop level, leading to antiproton
production from $Z$ boson decay. The corresponding Feynman diagrams are shown in the upper row of Fig.~\ref{fermion_loop}. We find for
the decay rate in the limit $m_\nn,m_\ell\ll m_\text{DM},M_Z$
\begin{eqnarray}\label{decay rate fdm to nn Z}
  \Gamma_{\fdm\to\nn Z^0} & = & \frac{3\alpha_{em}}{8\pi}\frac{m_{\rm DM}^5}{384(2\pi)^3}\left(1-\frac{M_Z^2}{m_{\rm DM}^2}\right)^2 \times F^\Sigma_Z(M_Z^2/m_{\rm DM}^2) \nonumber \\
  && \times\left|\sum_\ell \left(\frac{\lambda_{\ell N}^L \lambda_{\ell \psi}^L }{2 s_W c_W }\left(\frac{s_W^2-c_W^2}{m_{\Sigma_\pm}^2}+\frac{s_W^2+c_W^2}{m_{\Sigma_0}^2}\right) - \frac{\lambda_{\ell N}^R \lambda_{\ell \psi}^R s_W}{c_W m_{\sigma}^2}\right)\right|^2 \,,
\end{eqnarray}
where
\begin{equation}
  F^\Sigma_Z(x) = 1 + \frac{x}{2}\left(1 + 8\ln(m_\Sigma^2/M_Z^2) + \frac{16}{9}(1+2x)\left|\ln(m_\Sigma^2/M_Z^2)-i\pi\right|^2 \right)\;,
\end{equation}
with $m_\Sigma$ being of the order of the intermediate particle masses.\footnote{In general the masses appearing in the logarithms
in $F^\Sigma_Z$ can be different for each mediator. We assume here for simplicity that the masses are of similar size. Alternatively, if the rate is dominated by one intermediate particle, then $m_\Sigma$ corresponds to the mass of this field}

For example, for purely left-chiral couplings the branching ratios for loop-suppressed decays are given in this scenario by
\begin{equation}\label{ratio NZ Nll}
  \text{BR}(\fdm\to Z^0 N)
   =  \frac{3\alpha_\text{em}}{8\pi} \frac{|\sum_\ell \lambda_{\ell N}^L \lambda_{\ell \psi}^L  |^2}{\sum_\ell | \lambda_{\ell N}^L \lambda_{\ell \psi}^L  |^2} \big(1-\frac{M_Z^2}{m_{\rm DM}^2}\big)^2 F^\Sigma_Z 
\times \left\{ 
\begin{array}{ll} 
\frac{s_W^2}{2c_W^2} & m_{\Sigma_\pm}=m_{\Sigma_0} \\
\frac{(c_W^2-s_W^2)^2}{4s_W^2c_W^2} & m_{\Sigma_\pm}\ll m_{\Sigma_0} \\
\frac{(c_W^2+s_W^2)^2}{4s_W^2c_W^2} & m_{\Sigma_\pm}\gg m_{\Sigma_0} \\
\end{array}\right. \,,
\end{equation}
where the three cases refer to the possible mass orderings of the charged and neutral components
of the mediating scalar fields. For comparison, the branching ratio of the electromagnetic decay mode
for these cases is given by
\begin{equation}
  \text{BR}(\fdm\to \gamma N)
   =  \frac{3\alpha_\text{em}}{8\pi} \frac{|\sum_\ell \lambda_{\ell N}^L \lambda_{\ell \psi}^L  |^2}{\sum_\ell | \lambda_{\ell N}^L \lambda_{\ell \psi}^L  |^2} 
\times \left\{ 
\begin{array}{ll} 
\frac{1}{2} & m_{\Sigma_\pm}=m_{\Sigma_0} \\
1 & m_{\Sigma_\pm}\ll m_{\Sigma_0} \\
\frac{m_{\Sigma_0}^4}{m_{\Sigma_\pm}^4}\ll 1 & m_{\Sigma_\pm}\gg m_{\Sigma_0} \\
\end{array}\right. \,.
\end{equation}
As expected, both decay channels differ mainly due to the different coupling strength, as well as the different kinematics arising from the finite mass of the $Z$ boson. When $m_{\Sigma_\pm}=m_{\Sigma_0}$, the dark matter particle decays with
equal rate into leptons and neutrinos at tree level. Therefore, the branching ratio for the decay into photons
is half as large compared to the case $m_{\Sigma_\pm}\ll m_{\Sigma_0}$. In the case
$m_{\Sigma_\pm}\gg m_{\Sigma_0}$ the electromagnetic channel is relatively suppressed, while the radiative decay
producing $Z$ bosons is also possible via the vertex diagrams involving electrically neutral particles in the loop.

The diagrams  for the case when the particle mediating the decay is a vector boson are shown in the lower row of Fig.~\ref{fermion_loop}. The resulting
branching ratios can be obtained by replacing in Eq.~(\ref{decay rate fdm to nn Z})
$m_\Sigma\to m_V$, $3\alpha_\text{em}/(8\pi)\to 27\alpha_\text{em}/(8\pi)$ and
$F^\Sigma_Z\to F_Z^V$, where
\begin{equation}
  F_Z^V(x) = 1 + \frac{x}{2}\left(1 - \frac83 \left( \ln(m_V^2/M_Z^2) + 3 \right) + \frac{16}{81}(1+2x)\left|3+\ln(m_V^2/M_Z^2)-i\pi\right|^2 \right) \;.
\end{equation}

\begin{figure}[h!]
\begin{center}
\psfrag{.}{\scriptsize $. \, B_1$}
\psfrag{..}{\scriptsize $. \, B_2$}
\psfrag{...}{\scriptsize $. \, B_3$}
\psfrag{....}{\scriptsize $. \, B_4$}
\psfrag{.....}{\scriptsize $. \, B_5$}
\psfrag{......}{\scriptsize $. \, B_6$}
\psfrag{EZhhhhhhhhh}{\scriptsize $E_Z$ [GeV]}
\psfrag{mdmgammahhhhhhhhhhhhhhhhhhhhhhhhh}{\scriptsize \hspace{0.2cm} $m_\text{DM} \Gamma^{-1}(\psi_\text{DM} \to Z^0 N)$ [GeV s]}
\includegraphics[width=100mm]{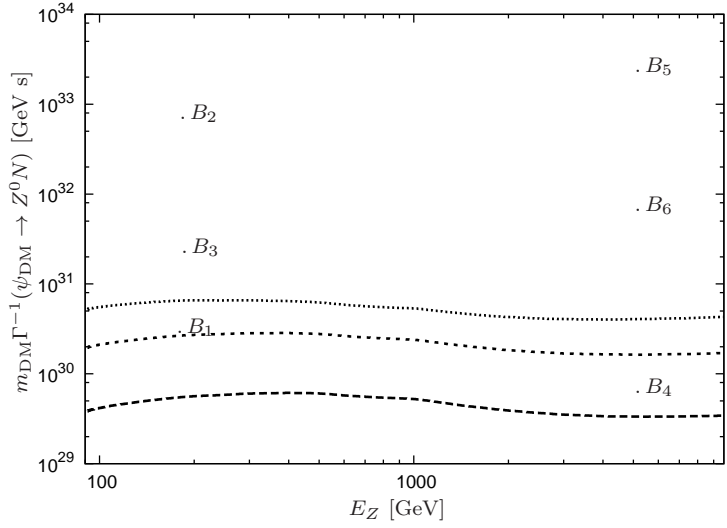}
\caption{Constraints on the decay $\psi_\text{DM} \to Z^0 N$ with $m_N > 0$. We show the benchmark points from Table~\ref{tab:benchmarks} for comparison.}
\label{fig:ZN}
\end{center}
\end{figure}

As for the case of the loop-induced decay $\psi\rightarrow \gamma N$, the rate for the process $\psi\rightarrow Z N$ can be enhanced in the degenerate limit. This is illustrated in Fig.~\ref{fig:ZN}, where we plot the constraints on $m_\text{DM} \Gamma^{-1}$ against the energy of the $Z$ boson. In the figure we also display the positions of the benchmark points from Table~\ref{tab:benchmarks} for comparison. We see that depending on the ratio between $m_{\text{DM}}$ and $m_N$, the radiative decays can vary in their relative intensity to lie either in the allowed or in the excluded region of the parameter space. Benchmark points 1 and 4, which correspond to the highest degeneracy and both have $\eta=-1$, can be ruled out based on the primary antiproton flux produced from the loop-induced decay into $Z$ bosons, when assuming MAX or MED propagation parameters, respectively. Points 3 and 6 also have an enhanced radiative decay, but are still in the allowed region. For comparison, points 2 and 5 (for which $\eta=+1$, so that there is no enhancement of the one-loop decay) are also shown. Note that the limits arising from the one-loop decay into monoenergetic gamma rays have been analyzed for the same set of benchmark points in~\cite{Garny:2010eg}. There, it was found that the photon flux predicted for benchmark points 1 and 4 is in conflict with Fermi LAT limits from the Milky Way halo, and MAGIC observations~\cite{Aleksic:2009ir} of the Perseus cluster, respectively. The photon flux of benchmark point 6 can be tested in the future by the CTA. Thus, we conclude that for dark matter that decays ``leptophilically'' at tree level, the limits from monochromatic gamma rays and from antiprotons produced in the one-loop decay channels into photons and $Z$ bosons, respectively, are competitive, although the former ones are slightly more restrictive in the case of MED and, especially, for MIN propagation parameters.

\begin{table}
  \begin{tabular}{cccccccc}
    \#&Channel & $\eta$ & $m_\text{DM}$ [GeV] & 
    $m_N$ [GeV] & $E_Z$ [GeV] & $\Gamma_{\ell^+\ell^-N}^{-1}$ [s] &
    $\frac{m_\text{DM}}{\Gamma_{\psi\to Z  N}}$ [s\ TeV] \\\hline
    1&$e_L^-e_L^+N$   &$-1$ &1000    &812.4 & 174 & $2.5\times10^{26}$
    &$3.30\times10^{27}$ 
    \\
    2&$e^-_Le^+_LN$   &$+1$ &500     &282.8 & 178 &$5\times10^{26}$
    &$7.98\times10^{29}$ \\
    3&$e^-_Le^+_LN$  &$-1$ &400    &154.9 & 180 & $6.3\times10^{26}$
    &$2.59\times10^{28}$ \\
    4&$\mu^-_L\mu^+_LN$ &$-1$ &100000  &94868 & 5000 & $4.5\times10^{24}$
    &$7.06\times10^{26}$ \\
    5&$\mu^-_L\mu^+_LN$ &$+1$ &15000   &8660  & 5000 &$3\times10^{25}$
    &$2.68\times10^{30}$\\
    6&$\mu^-_L\mu^+_LN$ &$-1$ &15000   &8660  & 5000 &$3\times10^{25}$
    &$7.49\times10^{28}$
  \end{tabular}
  \caption{Benchmark scenarios. In the first three cases, the three-body decay
  produces electrons only. In the last four cases, the three-body decay
  produces muons only.}
  \label{tab:benchmarks}
\end{table}

\subsection{The Decay $\phi_\text{DM} \rightarrow \gamma Z^0$}

Analogously to the fermionic case, we also consider a scalar dark matter particle that decays leptonically via its tree level coupling to charged leptons, $\sdm\to\ell^+\ell^-$. At the one-loop level, the dark matter can also decay into
a pair of gauge bosons. The rate for the loop-induced decay into a pair of photons, $\sdm\to\gamma\gamma$, has been discussed
in the context of gamma-ray line searches in Ref.~\cite{Garny:2010eg}. Antiprotons are produced in the loop-induced decays involving the massive gauge bosons.
The decay rate for the loop-induced process $\sdm\to\gamma Z^0$ can be adapted, \textit{e.g.}, from the results in 
Refs.~\cite{Bergstrom:1985hp,Spira:1991tj}. For the branching ratio, we find in the limit $m_\ell\ll m_\text{DM},M_Z$
\begin{align}
\text{BR}(\phi_\text{DM} \rightarrow \gamma Z^0) & = \frac{4\alpha_\text{em}^2 m_\ell^2(1-x) }{\pi^2 m_\text{DM}^2} \frac{(3s_W^2-c_W^2)^2}{s_W^2c_W^2}  \left|\frac{3}{4} - \ln(x) \left(\frac{1}{2} \left(\ln\left(x_\ell\right) - i\pi\right) + \frac{x}{1 - x}\right)\right|^2\nonumber\\
& \simeq 10^{-11} \left(\frac{m_\ell}{106~\text{MeV}}\right)^2 \left(\frac{1~\text{TeV}}{m_\text{DM}}\right)^2 \left(1 - \frac{M_Z^2}{m_\text{DM}^2}\right) \,,
\end{align}
where $x \equiv M_Z^2/m_\text{DM}^2$ and $x_\ell\equiv m_\sdm^2/m_\ell^2$. This radiative decay mode
is suppressed because of the necessary helicity flip of the lepton in the loop.
Analogously, the decays $\phi_\text{DM}\rightarrow Z^0 Z^0$ and $\phi_\text{DM} \rightarrow W^+ W^-$ are in principle allowed. We do not examine these decay modes in detail since they are suppressed in the same manner.

\section{Conclusions}
\label{sec:conclusions}
We have presented a general model-independent analysis of the cosmic-ray antiproton constraints on the partial dark matter lifetime in hadronic decay modes for the lowest-order decays allowed by Lorentz and gauge invariance. In particular, we have investigated the possible two-body decay modes for fermionic and scalar dark matter particles, as well as three-body decays into a pair of quarks and a neutral particle (for example, a neutrino) for Majorana dark matter particles. For the three-body decay we have paid special attention to the resulting energy spectra, which can differ slightly depending on whether the decay is mediated by a scalar or a vector particle, and whether the neutral particle produced in the final state is nearly massless or nearly degenerate with the dark matter particle. Employing a semi-analytical model of cosmic-ray propagation we have scanned the mass--lifetime parameter space over several orders of magnitude, and have derived lower limits on the lifetime for three representative sets of propagation parameters from the PAMELA measurement of the antiproton-to-proton ratio. We have also examined kinematic enhancement of radiative decays as well as the production of monochromatic gamma rays. Radiative decays into photons induced at next-to-leading order by tree-level decay into the electrically charged quarks provide an interesting complementarity to the antiproton constraints. Finally, we have discussed antiproton constraints arising from loop-induced decays into weak gauge bosons for leptophilic dark matter.

\section*{Acknowledgements}

This work has been partially supported by the DFG cluster of excellence ``Origin and Structure of the Universe'' (MG, AI, DT) and by the DFG Collaborative Research Center 676 ``Particles, Strings and the Early Universe'' (MG). DT also acknowledges support from the DFG Graduiertenkolleg ``Particle Physics at the Energy Frontier of New Phenomena.'' DT is grateful to Rolf Kappl and Martin Winkler for helpful discussions.

\section*{Note Added}
During the completion of this work an independent preprint \cite{Cheng:2012uk} appeared in which three-body dark matter decays and antiproton constraints are discussed.

\end{document}